\documentclass[11pt,a4paper]{article}

\usepackage[margin=2.5cm]{geometry}

\usepackage[utf8]{inputenc}
\usepackage[final]{microtype}
\usepackage{lmodern}
\usepackage{exscale}
\usepackage[T1]{fontenc}
\usepackage{microtype}

\usepackage{amsmath}
\usepackage{amssymb}
\usepackage{mathtools}

\numberwithin{equation}{section}

\usepackage{hyperref}
\usepackage[dvipsnames]{xcolor}
\hypersetup{
    colorlinks,
    linkcolor={purple},
    citecolor={purple},
    urlcolor={purple}
}

\usepackage{cite}

\usepackage{mathrsfs}

\usepackage{relsize,exscale}
\usepackage{tensor}

\usepackage[low-sup]{subdepth}

\newcommand{\scr}{\scriptscriptstyle}

\usepackage[low-sup]{subdepth}

\usepackage{graphicx}
\newcommand{\longsim}{\scalebox{1.8}[1]{$\sim$}}

\DeclareMathAlphabet{\mathbfi}{OML}{cmm}{b}{it}
\renewcommand{\vec}[1]{{\ifnum9<1#1\mathbf{#1}\else\ifcat\noexpand#1\relax\boldsymbol{#1}\else\mathbfi{#1}\fi\fi}}

\usepackage[font={small}]{caption}

\makeatletter
\newcommand{\dalembertian}{\mathop{\mathpalette\dalembertian@\relax}}
\newcommand{\dalembertian@}[2]{%
  \begingroup
  \sbox\z@{$\m@th#1\square$}%
  \dimen0=\fontdimen8
    \ifx#1\displaystyle\textfont\else
    \ifx#1\textstyle\textfont\else
    \ifx#1\scriptstyle\scriptfont\else
    \scriptscriptfont\fi\fi\fi3
  \makebox[\wd\z@]{%
    \hbox to \ht\z@{%
      \vrule width \dimen0
      \kern-\dimen0
      \vbox to \ht\z@{
        \hrule height \dimen0 width \ht\z@
        \vss
        \hrule height 2\dimen0
      }%
      \kern-2.5\dimen0
      \vrule width 2.5\dimen0
    }%
  }%
  \endgroup
}
\makeatother

\usepackage{cancel}

\usepackage[normalem]{ulem}

\usepackage{cancel}

\begin{document}

\begin{center}
{\bf \Large One-loop correction to primordial tensor modes \\ during radiation era}

\bigskip

\renewcommand{\thefootnote}{\fnsymbol{footnote}}

{Markus B. Fr{\"o}b},${}^a$\footnote{email: \href{mailto:mfroeb@itp.uni-leipzig.de}{\tt mfroeb@itp.uni-leipzig.de} }
{Dra\v{z}en Glavan},${}^b$\footnote{email: \href{mailto:glavan@fzu.cz}{\tt glavan@fzu.cz} }
{Paolo Meda}\,${}^{c,d}$\footnote{email: \href{mailto:paolo.meda@unitn.it}{\tt paolo.meda@unitn.it} }
{and Ignacy Sawicki}\,${}^{b}$\footnote{email: \href{mailto:sawicki@fzu.cz}{\tt sawicki@fzu.cz} }

\setcounter{footnote}{0} 

\bigskip

{${}^a$\,\it Institut f{\"u}r Theoretische Physik, Universit{\"a}t Leipzig, Br{\"u}derstra{\ss}e 16, 04103 Leipzig, Germany}

\medskip

{${}^b$\,\it CEICO, FZU --- Institute of Physics of the Czech Academy of Sciences,}
\\
{\it Na Slovance 1999/2, 182 21 Prague 8, Czech Republic}

\medskip

{${}^c$\,\it Dipartimento di Fisica, Universit{\`a} di Pavia, and Istituto Nazionale di Fisica Nucleare, 
\\
Gruppo IV, Via Bassi 6, 27100 Pavia, Italy}

\medskip

{${}^d$\,\it Dipartimento di Matematica, Universit{\`a} di Trento, 
\\
via Sommarive 14, I-38123 Povo (Trento), Italy}

\

\

\parbox{0.86\linewidth}{
The ability to infer properties of primordial inflation relies on the 
conservation of the superhorizon perturbations between their exit during 
inflation, and their re-entry during radiation era. Any considerable 
departure from this property would require reinterpreting the data. This is 
why it is important to understand how superhorizon perturbations interact with 
the thermal plasma driving the radiation dominated Universe. We
model the plasma by free photons in a thermal state and compute the one-loop
correction to the power spectrum of primordial tensor perturbations. 
This correction grows in time and is not suppressed by any small parameter. 
While one-loop result is not reliable because it invalidates perturbation theory, 
it signals potentially interesting effects that should be investigated further.
}

\end{center}

\

\section{Introduction}

The leading paradigm for the earliest stage of our Universe is primordial 
inflation, that posits our Universe began with a very brief period of very rapid 
accelerated expansion. Our current ability to infer precise properties of 
this period, and to constrain a plethora of proposed models~\cite{Martin:2013tda}
relies on interpreting how signals emitted much later are connected to processes
that took place during inflation. These signals come in the form of a cosmic microwave 
background (CMB) emitted some 380,000 years after the end of inflation.
The fluctuations in the CMB carry the imprints of density perturbations and
gravitational wave perturbations from the last scattering surface. These perturbations 
are supposed to have originated from primordial inflationary perturbations,
that have exited the horizon during inflation, with their amplitude frozen
until re-entering the horizon in late radiation-dominated period.
Therefore, our ability to infer the properties of primordial inflation hinges
on assuming the  conservation of primordial perturbations on superhorizon scales.
Any considerable departure from this property would alter the way we 
interpret data, and would necessitate revisiting the constraints on models 
of inflation delivered in~\cite{Planck:2018jri}.

The conservation of linearized superhorizon tensor modes indeed is valid for cosmological perturbations around backgrounds  the expansion of which is driven by homogeneous and isotropic condensates of matter fields~\cite{Mukhanov:1990me}. However, this has not been established 
for backgrounds driven by fluctuating media.  The thermal plasma driving the expansion of the universe during radiation domination is an instance of such a medium. Given that most observationally relevant modes re-enter the horizon only
by the end of radiation-dominated era, understanding the details of the evolution of superhorizon modes in a thermal medium is of paramount importance.

Only recently has this question  been considered more closely 
in~\cite{Ota:2023iyh}, with staggering conclusions: In the simplest 
one-loop approximation the primordial tensor power spectrum shows large secular 
enhancement, unsuppressed by a small parameter. We find this result to be of utmost 
importance as, taken at face value, it challenges our current ability to quantify the 
properties of the primordial Universe. That is why we set out to 
revisit independently the same question by considering 
a different system where the thermal radiation bath is composed out of 
photons, rather than excitations of a massless minimally coupled scalar as in~\cite{Ota:2023iyh}.

We find the same large secular one-loop enhancement as~\cite{Ota:2023iyh},
confirming the expectation that the effect does not depend
on the number of degrees of freedom in the plasma. This amplification has recently been reinterpreted 
in~\cite{Ota:2024idm} as the effect of stimulated emission of gravitational
waves. This enhancement is not suppressed by a small parameter naively 
expected at one loop, and signals  the breakdown of perturbative
loop expansion and/or the breakdown of linearity of evolution of superhorizon
primordial gravitational waves. While perturbative results obtained 
in~\cite{Ota:2023iyh}, and the ones we report here should not be trusted as predictions because of this breakdown,
they point to the pressing need to better understand the evolution
of long-wavelength primordial fluctuations during the radiation era when they propagate
on what are essentially stochastic backgrounds. The need for such an understanding 
has been advocated recently in~\cite{Liu:2024utl}, motivated by
the question of whether gravitational waves remain massless when propagating
on such backgrounds. Our findings conform with the tensor 
perturbations remaining massless, which at one loop level we find to be captured 
by the Hartree approximation.

The computation considered here pertains to the radiation-dominated era
following the post-inflationary reheating period, during which the 
expansion is dominated by a thermalized plasma of relativistic 
particle species. The standard assumptions about this period are
collected in Sec.~\ref{sec: Standard results for tensors in the early universe}.
We set up the computation in Sec.~\ref{sec: Setting up one-loop equation for GW}
starting from considering the Einstein and Maxwell equations
to describe the evolution of a priori stochastic variables. These equations 
are then turned into linearized equations for metric perturbations evolving 
on a background whose expansion is driven by thermal fluctuations. 
The interaction between metric perturbations and thermal fluctuations of the
plasma\footnote{We refer to the bath of noninteracting photons with 
the thermal spectrum as plasma for short.} are then truncated at the one-loop 
level when computing the two-point function for the tensor perturbations.
We find this approximation to correspond to the equations of
stochastic gravity~\cite{Hu:2008rga,Hu:2020luk} when linearized in metric 
perturbations.

The real-time, thermal, non-local  self-energy diagrams are computed in 
the vanishing external spatial momentum limit in 
Sec.~\ref{sec: Computing diagrams}, with the more complicated integrals
involved relegated to the Appendix. These are then used
to compute the connected diagrams contributing to the tensor power spectrum.
There are two qualitatively different contributions: (i) thermally induced gravitational wave production~\cite{Ghiglieri:2015nfa,Ghiglieri:2020mhm} 
that does not depend on the amplitude of the primordial perturbations
and is rather sourced by the plasma, and (ii) a
radiation exchange effect that takes the form
of a secular, multiplicative enhancement of the primordial tensor power 
spectrum. While we find a different result compared to~\cite{Ota:2023iyh}
for the first blue-tilted contribution, this is of little consequence as 
it is negligible for the observationally relevant scales.
It is only the second effect that is relevant, for which we agree 
with~\cite{Ota:2023iyh}.
Given that this effect is captured by a linear equation,
it is possible to solve for its evolution exactly, and thus to resum 
the infinite series of self-energy insertions. We do this in 
Sec.~\ref{sec: Resummed computation}
to reveal that the growing logarithmic secular correction becomes a 
power-law growing secular enhancement. Such a rapid growth indicates
the breakdown of the loop expansion
and/or the breakdown of the linear
evolution of primordial tensor perturbations on superhorizon scales.
The relevance of these results is discussed in the concluding 
section, where we also outline what we think should be the next steps in 
understanding this problem.

\section{Standard results for tensors in the early universe}
\label{sec: Standard results for tensors in the early universe}

One of the key predictions of inflationary cosmology, 
though as of yet unobserved, is the existence of a stochastic background of
primordial gravitational waves. These are defined as perturbations~$h_{\mu\nu}$
of the conformally rescaled Friedmann-Lemaître-Robertson-Walker (FLRW) metric,
\begin{equation}
g_{\mu\nu} = a(\eta)^2 \bigl( \eta_{\mu\nu} 
    +
    \kappa h_{\mu\nu} \bigr)
\, ,
\label{MatricPerturbation}
\end{equation}
where~$\eta_{\mu\nu}$ is the Minkowski metric,~$\eta$ is the conformal 
time,~$a(\eta)$ is the scale factor, and where we find convenient to factor 
out~$\kappa \!=\! \sqrt{16\pi G_{\scr \rm N}}\!=\! \sqrt{2}/ M_{\scr \rm P}$
from the definition of the fluctuation. The two propagating tensor polarisations are 
contained in the transverse traceless (TT) part of the metric fluctuations and we 
will find it convenient to use as our variable
\begin{equation}
\gamma_{ij} = (h_{ij})^{\scr \rm TT} \ ,
\end{equation}
so that~$\gamma_{ii} \!=\! 0$, and~$\partial_i \gamma_{ij} \!=\! 0$. 
They evolve according to the equation,
%
%
%
\begin{equation}
\Bigl( \partial_0^2 
    + 2 \mathcal{H} \partial_0
    - \nabla^2 \Bigr) \gamma_{ij}
    = 0 \, ,
\label{eq:h-eom}
\end{equation}
where~$\partial_0$ denotes a derivative with respect to conformal time,
where~$\mathcal{H}\!=\!\partial_0 a/a$ is the conformal Hubble rate related to the
physical Hubble rate as~$\mathcal{H}\!=\!aH$, and 
where~$\nabla^2\!=\!\partial_i \partial_i$ is the Laplacian.

The initially small subhorizon fluctuations are amplified by the inflationary 
background as their wavelenghts stretch to superhorizon scales. 
This property of tensor fluctuations is described by the 
coincident limit of the two-point correlation function expressed in momentum space,
\begin{equation}
    \bigl\langle \gamma_{ij}(\eta, \vec{x} ) 
        \gamma_{\scr ij}(\eta, \vec{x}' ) \bigr\rangle 
    = \int\! \frac{d^3\vec{k}}{(2\pi)^3} \, 
        e^{i\vec{k}\cdot(\vec{x}-\vec{x}')} 
        P_t^0(\eta,k)
        \, ,
\end{equation}
where we have already assumed spatial homogeneity and isotropy of the background during inflation,
so that the dimensionful power spectrum~$P_t$ is a function of momentum modulus
only,~$k\!=\!\| \vec{k} \|$. The dimensionful power spectrum is often expressed 
in terms of its dimensionless counterpart,
\begin{equation}
\mathcal{P}_t^0(\eta,k) = \frac{\kappa^2 k^3}{2\pi^2} P_t^0(\eta,k) \, ,
\end{equation}
that is the main object of our discussion. The inflationary spacetime 
is a quasi-de Sitter period during 
which~$\mathcal{H}\!=\!-1/(H_{\rm inf} \eta )$,
with~$\eta$ negative, and $H_\text{inf}$ the nearly constant physical 
Hubble parameter. Normalising the modes to the 
Chernikov-Tagirov-Bunch-Davies~\cite{Chernikov:1968zm,Bunch:1978yq}
vacuum deep inside the cosmological horizon, 
one obtains as the solution a key prediction of inflation: 
the power spectrum obtains an amplitude given by the curvature scale of 
the inflating background at the moment the mode 
crosses the cosmological horizon~\cite{Grishchuk:1975uf, Starobinsky:1979ty, Rubakov:1982df, Abbott:1984fp} (for a now-textbook presentation see \cite{Lyth:2009zz}),

\begin{equation}
    \mathcal{P}^0_t
	= \frac{\kappa^2H_\text{inf}^2}{\pi^2} \biggr|_{k=\mathcal{H}}
	\, .
\end{equation}
Since the expansion is accelerating during inflation, these modes 
continue being stretched, but their amplitude remains constant once they have 
crossed the horizon. Inflation thus predicts a primordial tensor power spectrum 
which is homogeneous, isotropic, with the only deviation from scale invariance 
resulting from the slow evolution of $H_\text{inf}$ 
during the quasi-de Sitter phase. This is usually described by the tensor tilt,
\begin{equation}
n_t \equiv \frac{ d \ln \mathcal{P}^0_t}{ d \ln k} \,.
\end{equation}
Rather than giving the amplitude of tensor directly (since it is not yet observed), one usually 
constrains the ratio $r$ of the amplitude of the power spectrum of tensor fluctuations to the 
already observed scalar fluctuations. In the simplest single-scalar field models of inflation the 
tensor tilt is simply related to $r$ through, $n_t \!=\! r/8$, although this relation is modified in 
alternative models.

\medskip

Eventually, the slow-roll phase ends, the universe reheats and the radiation-domination 
era begins. At this point the universe is filled with an extremely hot and interacting plasma 
with temperature~$T_\text{rh}$, while the Hubble parameter is given by
\begin{equation}
H_\text{rh}^2 = \frac{\pi^2\kappa^2}{180}g_\text{eff}T^4_\text{rh} \,,
\label{Fri}
\end{equation}
where $g_\text{eff}$ is the effective number of relativistic degrees of freedom 
fixed by the high-energy physics at the reheating temperature. The temperature then 
scales as~$a^{-1}$ giving $\mathcal{H} \!=\! H_\text{rh} (a_{\rm rh}/a)$ --- the 
expansion decelerates. The equation of motion for the tensor fluctuations~\eqref{eq:h-eom} 
then has as the solution in the superhorizon limit, $k \! \ll\! \mathcal{H}$, a constant mode 
and a decaying one. This means that during radiation domination, the evolution of the 
tensor fluctuations is effectively frozen on superhorizon scales. In the concordance 
scenario, there is no evolution of the amplitude superhorizon between reheating and 
when the mode shrinks sufficiently to re-enter the cosmological horizon.

This simple picture is slightly modified by the damping of the amplitude squared of gravitational waves by around one third by neutrinos when their wavelengths are comparable to the cosmological horizon, first described in Ref.~\cite{Weinberg:2003ur}. Such damping was later investigated much more generally in Ref.~\cite{Baym:2017xvh}, showing that no significant additional modification of gravitational wave amplitude can be caused by the matter sources of the type present in the universe in approximately equilibrium distributions. Other sources of gravitational waves, such as topological defects or black-hole inspirals, are only active at subhorizon scales and therefore do not contribute to the horizon-scale observations in the CMB.

The tensor fluctuations that re-enter the horizon close to the time of matter-radiation 
equality contribute to the power spectrum of temperature fluctuations in the CMB at the 
largest scales, and are a unique source of primary B-modes in the polarisation power spectrum. Since the details of the radiation era are completely transparent to the dynamics 
of superhorizon tensor fluctuations (and the scalar ones for similar reasons), 
the Einstein-Boltzmann codes used for predicting the power spectra of fluctuations 
at the last scattering surface (e.g.~CAMB \cite{Lewis:1999bs} and CLASS \cite{Lesgourgues:2011re}) , do not consider the bulk of 
the radiation era at all, but rather start the modes' evolution when their wavelengths 
are some order of magnitude larger than the Hubble scale. A measurement of the tensor 
amplitude from the last cosmic microwave background is thus really a constraint on 
the amplitude as it was close to the time of matter-radiation equality. We are 
belabouring this point, since the main result of this paper is the statement that 
there is a superhorizon correction to the evolution equation~\eqref{eq:h-eom} during 
the radiation domination era which challenges the superhorizon conservation law.
\medskip

Let us fix some orders of magnitude and connect to parameters being constrained 
by observations. The headline Planck constraint {~\cite{Planck:2018jri} on tensor 
modes, $r_\text{0.002}<0.010$, comes from assuming the single-scalar-field prediction for the tensor tilt and a pivot in $r$ at the scale $k=0.002~\text{Mpc}^{-1}$ and is obtained from the temperature power spectrum. Alternative analyses leave the tensor tilt free (although constant), constraining the tensor modes at a second 
scale $k=0.02~\text{Mpc}^{-1}$, giving $r_{0.002}<0.044$ and $r_{0.02}<0.184$. 
Finally, BICEP/KECK~\cite{BICEP:2021xfz}
constrains the polarisation B-modes directly 
at the pivot scale $k=0.05~\text{Mpc}^{-1}$ assuming $n_t=0$, giving $r<0.036$.
For us, the most relevant information is that the smallest of these scales re-enters the 
horizon when non-relativistic matter provides a significant 10\% of the energy budget of 
the universe, with this fraction larger for the longer modes. As a first-order approximation, 
we will assume that all these modes are affected by the physics presented in this paper 
equally and will not discuss the potential additional scale dependence induced at the end 
of the radiation-domination epoch.

\medskip

Of key importance in our computation is the duration of the radiation domination era. 
This is not known, beyond the fact that the Universe must have reheated to a temperature 
high enough to allow for Big Bang Nucleosynthesis,~$T_{\rm \scr BBN} \!\sim\! 1\,\text{MeV}$. 
Assuming particular classes of inflation models, one can obtain constraints on the reheating 
temperature from Planck observations~\cite{Martin:2014nya}, but they are not strong and the range of possibilities is wide; for example~$T_\text{rh} \!>\! 400\,\text{TeV}$ 
for small-field models, $T_\text{rh} \!>\! 18\,\text{PeV}$ for supergravity scenarios, but $T_\text{rh} \!\sim\! \mathcal{O}(10^{13} \, {\rm GeV})$ \cite{Planck:2018jri} for Higgs inflation \cite{Bezrukov:2007ep}. 
Reheating is thought to be short and to occur soon after inflation ends (see e.g.~Ref.~\cite{Allahverdi:2010xz} for a review of reheating mechanisms), although for some mechanisms this can be extended, adding to the uncertainly (e.g.~reheating through primordial black hole evaporation \cite{Gross:2024wkl}). We take as a fiducial temperature scale of reheating ~$T_\text{rh} \!\sim\! 10^{13}\,\text{GeV}$. We wish to 
compute the ratio of the scale factor at reheating $a_\text{rh}$ and at the end of the 
radiation-domination era $a_*$. This is given by
\begin{equation}
    \exp(\mathcal{N}_\text{rad}) = \frac{a_*}{a_\text{rh}} = \left(\frac{g_S(T_\text{rh})}{g_S(T_0)}\right)^{1/3} \frac{T_\text{rh}}{T_0} a_*
\end{equation}
with $T_0$ the CMB temperature today, and $g_{\scr S}$ the effective number 
of entropy degrees of freedom, with $g_{\scr S}(T_0) \!=\! 3.91$ 
while the value at reheating depends on the particle 
content of the universe at that energy scale. Usually one 
takes~$g_{\scr S}(100\,\text{GeV}) \!\sim\! 100$ 
(see e.g.~\cite{Caprini:2018mtu}), 
but at the large energy scales of reheating this is unknown. Nonetheless, unless it 
is very large, it does not significantly contribute to $\mathcal{N}_\text{rad}$. For a fiducial duration of the radiation era we thus have,
\begin{equation}
    \mathcal{N}_\text{rad}
    =
    52 + \ln \left(\frac{T_\text{rh}}{10^{13}~\text{GeV}}\right) + \frac{1}{3}\ln \left(\frac{g_S(T_\text{rh})}{100}\right) \,.
\end{equation}
This is long enough that even logarithmic corrections to the standard expectations of superhorizon conservation could be a significant correction. It what follows we extend the standard formalism to account for the fluctuating nature of the radiation-era background.

\section{Setting up one-loop equation for tensor perturbations}
\label{sec: Setting up one-loop equation for GW}

The expansion during the radiation-dominated period of cosmology is driven by
the thermal plasma, rather than classical condensates of matter fields as
during inflation. For that reason we should consider the matter fields 
sourcing the Einstein equation as stochastic random variables. For consistency,
the left-hand side of the Einstein equation should have a stochastic character
as well, and thus the metric should be considered to be stochastic as well.
This is why we start by considering the radiation dominated period to be 
described by the Einstein and Maxwell equations,
\begin{equation}
G_{\mu\nu} = \frac{\kappa^2}{2} T_{\mu\nu} \, ,
\qquad \quad
\nabla^\mu F_{\mu\nu} = 0 \, ,
\label{EOMs}
\end{equation}
with the photon energy-momentum tensor,
\begin{equation}
T_{\mu\nu} = \Bigl( \delta_\mu^\rho \delta_\nu^\sigma
	- \frac{1}{4} g_{\mu\nu} g^{\rho\sigma}
	 \Bigr) g^{\alpha\beta} F_{\rho\alpha} F_{\sigma\beta}
	 \, ,
\end{equation}
where all the quantities are in principle stochastic random variables. We
assume for simplicity that the plasma driving the expansion can be modeled
by free-streaming photons with a thermal distribution. Starting from these
equations we shall derive one-loop equations for the metric perturbations
around homogeneous and isotropic backgrounds. In particular, in the following
we derive the equation of motion for the two-point function of traceless
transverse perturbations.

\subsection{Perturbing equations}

The assumption of the setup is that the matter sector has thermalized 
efficiently during the reheating period, without disrupting the homogeneity
and isotropy of the gravitational background. That is why we treat the
stochastic photon field as the background, while treating the stochastic
feature of the metric perturbatively. Consequently we want to derive a 
linear equation for the metric perturbation~$h_{\mu\nu}$,
defined in~(\ref{MatricPerturbation}) as the perturbation of the conformally
rescaled FLRW metric.
This metric perturbation induces perturbations of all the quantities appearing 
in equations of motion~\eqref{EOMs}, and we expand to linear order,
\begin{equation}
G_{\mu\nu} = G_{\mu\nu}^{\scr (0)} + \kappa G_{\mu\nu}^{\scr (1)} \, ,
\qquad
T_{\mu\nu} = T_{\mu\nu}^{\scr (0)} + \kappa T_{\mu\nu}^{\scr (1)} \, ,
\qquad
F_{\mu\nu} = F_{\mu\nu}^{\scr (0)} + \kappa F_{\mu\nu}^{\scr (1)} \, .
\end{equation}
Note that the power of~$\kappa$ is explicitly factored out of the perturbed
quantities, and that the parenthesized superscripts denote the number of
metric perturbations~$h_{\mu\nu}$ the quantity contains.

The background part of the Einstein tensor is just
\begin{equation}
G_{\mu\nu}^{\scr (0)}
    =
    \Bigl(
    3 \delta_\mu^0 \delta_\nu^0
    +
    (2\epsilon \!-\! 3) 
    \overline{\eta}_{\mu\nu} 
    \Bigr)
    \mathcal{H}^2
    \, ,
\label{zerothG}
\end{equation}
where~$\overline{\eta} \!=\! \eta_{\mu\nu} \!+\! \delta_\mu^0 \delta_\nu^0 $,
and where,~$\epsilon \!=\! 1 \!-\! \mathcal{H}'/\mathcal{H}^2 \!=\! - \dot{H}/H^2 $
is the so-called principal slow-roll parameter 
(in radiation-dominated period~$\epsilon \!=\! 2$). 
The perturbation of the Einstein tensor reads
\begin{align}
G_{\mu\nu}^{\scr (1)}
    ={}&
    - \frac{1}{2} 
    \Bigl[
    \partial^2 h_{\mu\nu}
    -
    2 \partial^\rho \partial_{(\mu} h_{\nu)\rho}
    +
    \partial_\mu \partial_\nu h
    +
    \eta_{\mu\nu} \bigl( \partial^\rho \partial^\sigma h_{\rho\sigma}
            - \partial^2 h \bigr)
    \Bigr]
\nonumber \\
&
    - \mathcal{H} \Bigl[
    2 \partial_{(\mu} h_{\nu)0}
    -
    \partial_0 h_{\mu\nu}
    -
    \eta_{\mu\nu} \bigl( 2 \partial^\rho h_{\rho 0} 
        - \partial_0 h \bigr)
    \Bigr]
    +
    (2\epsilon \!-\! 3) \mathcal{H}^2 
        \bigl( h_{\mu\nu} + \eta_{\mu\nu} h_{00} \bigr)
    \, ,
\label{EinsteinPerturbation}
\end{align}
where $h \!=\! h^\mu{}_\mu$ denotes the trace,
and where $\partial^2 \!=\! \partial^\mu \partial_\mu 
\!=\! - \partial_0^2 \!+\! \nabla^2$ is the flat space d'Alembertian
operator.
The background energy-momentum tensor matches the flat space expression on
account of conformal coupling of photons to gravity,
\begin{equation}
T_{\mu\nu}^{\scr (0)} =
    \frac{1}{2a^2}
    \Bigl( \delta_{(\mu}^\rho \delta_{\nu)}^\sigma
	- \frac{1}{4} \eta_{\mu\nu} \eta^{\rho\sigma}
	 \Bigr) \eta^{\alpha\beta}
     \bigl\{ F_{\rho\alpha}^{\scr(0)} ,
        F_{\sigma\beta}^{\scr (0)} \bigr\}
    \, ,
\end{equation}
where $\{ A, B \} \!=\! A B \!+\! B A$ denotes the anticommutator.
The perturbation of the energy-momentum tensor contains two contributions: 
(i) perturbation of the explicit metric dependence, and (ii) implicit 
perturbation contained in the perturbation of the photon field strength tensor,
\begin{equation}
T_{\mu\nu}^{\scr (1)} =
    \frac{1}{2a^2}
    {W_{\mu\nu}}^{\rho\sigma\alpha\beta\gamma\delta}
    \bigl\{ F_{\rho\alpha}^{\scr (0)} ,
        F_{\sigma\beta}^{\scr (0)} \bigr\}
        h_{\gamma\delta}
    +
    \frac{1}{a^2} \Bigl( \delta_\mu^\rho \delta_\nu^\sigma
        - \frac{1}{4} \eta_{\mu\nu} \eta^{\rho\sigma}
        \Bigr) \eta^{\alpha\beta} 
        \bigl\{ F_{\rho\alpha}^{\scr (0)} , F_{\sigma\beta}^{\scr (1)} \bigr\}
	 \, ,
\end{equation}
where the tensor structure is
\begin{equation}
{W_{\mu\nu}}^{\rho\sigma\alpha\beta\gamma\delta}
    =
    - \delta_{(\mu}^\rho \delta_{\nu)}^\sigma 
        \eta^{\gamma(\alpha} \eta^{\beta)\delta}
        + \frac{1}{2} \eta_{\mu\nu} \eta^{\rho\sigma} 
            \eta^{\gamma(\alpha} \eta^{\beta)\delta}
        -
        \frac{1}{4}
            \delta_{(\mu}^\gamma \delta_{\nu)}^\delta
            \eta^{\rho\sigma} \eta^{\alpha\beta}
            \, .
\label{Wtensor}
\end{equation}

The Maxwell equation allows us to express the implicit perturbation explicitly
in terms of the metric perturbation. The background stochastic photon
field strength satisfies the flat space 
equation,~$\partial_\nu F^{\nu\mu}_{\scr (1)}\!=\!0$, while its first
perturbation satisfies the equation sourced by the metric perturbation,
\begin{equation}
\partial_\nu F^{\nu\mu}_{\scr (1)}
    =
    \frac{1}{2} V^{\mu\rho\nu\sigma\alpha\beta}
    \partial_\nu \bigl( F_{\rho\sigma}^{\scr (0)} h_{\alpha\beta} \bigr)
    \equiv J^\mu_{\scr (1)}
    \, ,
\label{MaxwellFirst}
\end{equation}
where the tensor structure is
\begin{equation}
V^{\mu\rho\nu\sigma\alpha\beta}
    =
    4 \eta^{\alpha)[\mu} \eta^{\nu][\rho} \eta^{\sigma](\beta}
    +
    \eta^{\mu[\rho} \eta^{\sigma]\nu} \eta^{\alpha\beta}
    \, .
\label{Vtensor}
\end{equation}
The source of Eq.~(\ref{MaxwellFirst}) is necessarily 
conserved,~$\partial_\mu J^\mu_{\scr (1)} = 0$. With the help of Bianchi
identity the solution of this equation is found to be
\begin{equation}
F_{\mu\nu}^{\scr (1)}(x) = 
    \int\! d^4x' \, G_{\scr \rm R}(x \!-\! x') 
    \times 2 \partial'_{[\mu} J_{\nu]}^{\scr (1)}(x') \, ,
\end{equation}
where $G_{\scr \rm R}$ is the retarded Green's function that satisfies the equation
\begin{equation}
\partial^2 G_{\scr \rm R}(x \!-\! x') = \delta^4(x \!-\! x') \, .
\end{equation}
This retarded Green's function is conveniently expressed in terms of its spatial Fourier transform,
\begin{equation}
G_{\scr \rm R}(x \!-\! x')
    =
    \int\! \frac{d^3\vec{k} }{(2\pi)^3} \, 
    e^{i \vec{k} \cdot (\vec{x} - \vec{x}')}
    \widetilde{G}_{\scr \rm R}\bigl( \eta \!-\! \eta' \big| k \bigr),
\end{equation}
where~$k\!=\!\| \vec{k} \|$ is the momentum vector modulus, and where the
momentum space retarded Green's function reads
\begin{equation}
\widetilde{G}_{\scr \rm R} \bigl( \eta \!-\! \eta' \big| k \bigr)
    =
    - \theta(\eta \!-\! \eta') 
    \frac{ \sin\bigl[ k(\eta \!-\! \eta') \bigr] }{k}
    \, .
\end{equation}
Therefore, the perturbation of the photon field strength tensor is found to be
\begin{equation}
F_{\mu\nu}^{\scr (1)}(x) = 
    \int\! d^4x' \, G_{\scr \rm R}(x \!-\! x')  \,
    \partial'_\lambda \partial'_{[\mu}
    \Bigl[ 
        {V_{\nu]}}^{\rho\lambda\sigma\alpha\beta}
        F_{\rho\sigma}^{\scr (0)} (x') 
        h_{\alpha\beta}(x') \Bigr]
    \, .
\end{equation}
Implicit in the expression above is the lower limit of temporal integration
that is assumed to be the reheating time~$\eta_{\rm rh}$. Thus the photon
field strength perturbation vanishes at that time, which accounts for
the assumption of rapidly thermalized plasma during reheating in a process
not much influenced by gravitational interactions.

Having solved for the photon field strength perturbation, we can express 
the perturbed Einstein equation solely in terms of the metric perturbation.
The energy-momentum tensor now takes a nonlocal form,
\begin{align}
\MoveEqLeft[5]
T_{\mu\nu}^{\scr (1)}(x)
    =
    \frac{1}{2a^2}
    {W_{\mu\nu}}^{\rho\sigma\alpha\beta\gamma\delta}
    \bigl\{ F_{\rho\alpha}^{\scr (0)}(x) ,
        F_{\sigma\beta}^{\scr (0)}(x) \bigr\}
        h_{\gamma\delta}(x)
    +
    \frac{1}{a^2} \Bigl( \delta_{(\mu}^\rho \delta_{\nu)}^\sigma
        - \frac{1}{4} \eta_{\mu\nu} \eta^{\rho\sigma}
        \Bigr) \eta^{\alpha\beta} 
\nonumber \\
&
    \times \!
	\int\! d^4x' \, G_{\scr \rm R}(x \!-\! x')  \,
    \partial'_\kappa \partial'_{[\sigma}
    \Bigl[ 
        {V_{\beta]}}^{\theta\kappa\lambda\gamma\delta}
        \bigl\{ F_{\rho\alpha}^{\scr (0)}(x) , 
            F_{\theta\lambda}^{\scr (0)} (x') \bigr\}
        h_{\gamma\delta}(x') \Bigr]
        \, ,
\end{align}
owing to solving the Maxwell equation. The Einstein equation is now
written in terms of the thermally fluctuating~$F_{\mu\nu}^{\scr (0)}$
on an FLRW background characterized by the scale factor~$a^2$, and linearized in 
the metric perturbation~$h_{\mu\nu}$.

We proceed by introducing an additional assumptions about the system,
namely that the background metric is driven by the expectation value of the
background energy-momentum tensor,
\begin{equation}
G_{\mu\nu}^{\scr (0)} = \frac{\kappa^2}{2} \bigl\langle T_{\mu\nu}^{\scr (0)} \bigr\rangle \, ,
\label{BackgroundEinsteinEq}
\end{equation}
which is given by
\begin{equation}
\bigl\langle T_{\mu\nu}^{\scr (0)} \bigr\rangle =
    \frac{1}{2a^2}
    \Bigl( \delta_{(\mu}^\rho \delta_{\nu)}^\sigma
	- \frac{1}{4} \eta_{\mu\nu} \eta^{\rho\sigma}
	 \Bigr) \eta^{\alpha\beta} \bigl\langle \bigl\{
     F_{\rho\alpha}^{\scr(0)} , F_{\sigma\beta}^{\scr (0)} 
     \bigr\} \bigr\rangle
    \, .
\label{T0expectationFormal}
\end{equation}
This corresponds to the assumption that the thermal plasma is the dominant 
matter component during radiation-dominated period.
This allows to write the linearized Einstein equation as
\begin{equation}
G_{\mu\nu}^{\scr (1)}
    =
    \frac{\kappa}{2} 
    \Bigl( T_{\mu\nu}^{\scr (0)} - \bigl\langle T_{\mu\nu}^{\scr (0)} \bigr\rangle \Bigr)
    +
    \frac{\kappa^2}{2} T_{\mu\nu}^{\scr (1)}
    \, .
\label{FirstOrderEOM}
\end{equation}
We make another reorganization of this linearized equation by subtracting
the Hartree term from both sides of the equation, so that we write it in
the form
\begin{equation}
G_{\mu\nu}^{\scr (1)} 
    -
    \frac{\kappa^2}{4a^2}
    {W_{\mu\nu}}^{\rho\sigma\alpha\beta\gamma\delta}
    \bigl\langle \bigl\{ F_{\rho\alpha}^{\scr (0)} ,
        F_{\sigma\beta}^{\scr (0)} \bigr\} \bigr\rangle
        h_{\gamma\delta}
    =
    \frac{S_{\mu\nu}}{2a^2}
    \, ,
\label{HartreeRemovedEOM}
\end{equation}
where we split the right-hand side into three 
pieces,~$S_{\mu\nu} \!=\! S_{\mu\nu}^{\scr \rm I} 
    \!+\! S_{\mu\nu}^{\scr \rm II} 
    \!+\! S_{\mu\nu}^{\scr \rm III}$.
The first of these pieces does not depend on the metric perturbation,
but rather on the fluctuation of the background energy-momentum tensor,
\begin{equation}
S_{\mu\nu}^{\scr \rm I}(x) = 
    \kappa a^2
    \Bigl( T_{\mu\nu}^{\scr (0)}(x) 
        - \bigl\langle T_{\mu\nu}^{\scr (0)}(x) \bigr\rangle \Bigr)
        \, .
\end{equation}
The other two sources depend on the metric perturbation: 
the second one locally,
\begin{equation}
S_{\mu\nu}^{\scr \rm II}(x)
    =
    \frac{\kappa^2}{2}
    {W_{\mu\nu}}^{\rho\sigma\alpha\beta\gamma\delta}
    \Bigl( 
        \bigl\{ F_{\rho\alpha}^{\scr (0)}(x) , 
            F_{\sigma\beta}^{\scr (0)}(x) \bigr\}
        -
        \bigl\langle \bigl\{ F_{\rho\alpha}^{\scr (0)}(x) ,
            F_{\sigma\beta}^{\scr (0)}(x) \bigr\} \bigr\rangle
        \Bigr)
        h_{\gamma\delta}(x)
        \, ,
\end{equation}
and the third one nonlocally,
\begin{align}
S_{\mu\nu}^{\scr \rm III}(x)
    ={}&
    \kappa^2
    \Bigl( \delta_{(\mu}^\rho \delta_{\nu)}^\sigma
	- \frac{1}{4} \eta_{\mu\nu} \eta^{\rho\sigma}
	\Bigr) \eta^{\alpha\beta} 
\nonumber
\\
&   
    \times \!
    \int\! d^4x' \, G_{\scr \rm R}(x \!-\! x')  \,
    \partial'_\kappa \partial'_{[\sigma}
    \Bigl[ 
        {V_{\beta]}}^{\theta\kappa\lambda\gamma\delta}
        \bigl\{ F_{\rho\alpha}^{\scr (0)}(x) , 
            F_{\theta\lambda}^{\scr (0)} (x') \bigr\}
        h_{\gamma\delta}(x') \Bigr]
    \, .
\end{align}
This step of subtracting the Hartree term will ensure that the transverse
traceless component of the metric perturbation remains massless when
propagating on the fluctuating background.

We note that the background equation of motion~\eqref{BackgroundEinsteinEq} and the first-order one ~\eqref{FirstOrderEOM} can alternatively be derived in the stochastic gravity formalism~\cite{Hu:2008rga,Hu:2020luk}. The stochastic Langevin equation that is considered there reads~\footnote{Note that in the general stochastic gravity equation also tensors that arise from (the variation of) finite parts of counterterms appear. Since the renormalization of perturbative quantum gravity as an effective field theory is well understood (e.g.~\cite{Burgess:2003jk}), and the corresponding terms do not contribute to the leading corrections for large temperature, we may ignore them for our purposes. In general those tensors result in higher-derivative terms, which can pose problems for the solution of the equations~\cite{Meda:2020smb}. However, if those higher
derivative terms are treated perturbatively, in the same spirit in which they arise, the problems also disappear (e.g.~\cite{Glavan:2017srd,Glavan:2024cfs}).}
\begin{equation}
G_{\mu\nu}[g \!+\! \kappa h] 
    = \frac{\kappa^2}{2} 
        \Bigl( \bigl\langle T_{\mu\nu}[g \!+\! \kappa h] \bigr\rangle_{\scr A} + \xi_{\mu\nu}[g] \Bigr) \, ,
\label{Langevin}
\end{equation}
where $\langle \cdot \rangle_{\scr A}$ denotes the expectation value over the vector matter field only. Linearizing this equation in the metric perturbation $h_{\mu\nu}$, and considering the stochastic field $\xi_{\mu\nu}$ to be of the same order as $h_{\mu\nu}$, we obtain Eq.~\eqref{BackgroundEinsteinEq} at zeroth order and
\begin{equation}
G_{\mu\nu}^{\scr (1)} = \frac{\kappa^2}{2} \bigl\langle T_{\mu\nu}^{\scr (1)} \bigr\rangle_{\scr A} + \frac{\kappa}{2} \xi_{\mu\nu}[g]
\end{equation}
at first order. Since $\xi_{\mu\nu}$ is a stochastic variable with vanishing mean $\bigl\langle \xi_{\mu\nu}[g] \bigr\rangle \!=\! 0$ and 
covariance $\bigl\langle \xi_{\mu\nu}(x) \xi_{\rho\sigma}(x') \bigr\rangle \!=\! \left\langle \bigl[ T_{\mu\nu}^{\scr (0)}(x) - \bigl\langle T_{\mu\nu}^{\scr (0)}(x) \bigr\rangle \bigr]
    \bigl[ T_{\rho\sigma}^{\scr (0)}(x') - \bigl\langle T_{\rho\sigma}^{\scr (0)}(x') \bigr\rangle \bigr] \right\rangle$, 
this is exactly equivalent to Eq.~\eqref{FirstOrderEOM}.

\subsection{Vector field strength correlator}

The photon field does not see directly the expansion of the cosmological
background due to its conformal coupling to gravity. Therefore, its
thermal correlator takes the same form as in flat space,
\begin{equation}
\bigl\langle \bigl\{
	F_{\mu\nu}^{\scr (0)}(x) , F_{\rho\sigma}^{\scr (0)}(x')
	\bigr\} \bigr\rangle
	=
	8 \partial_{[\mu} \eta_{\nu][\sigma} \partial'_{\rho]}
	F(x \!-\! x')
	\, ,
\label{PhotonStatistical}
\end{equation}
where the thermal statistical two-point function~\cite{Bellac:2011kqa}
of the scalar field in flat space
\begin{equation}
F(x \!-\! x')
	=
	\int \!
	\frac{d^3 \vec{k}}{(2 \pi)^3} \,
	e^{ i \vec{k} \cdot (\vec{x} - \vec{x}') }
	\widetilde{F}\bigl( \eta \!-\! \eta' \big| k \bigr)
\end{equation}
is most conveniently expressed in terms of its spatial Fourier transform
\begin{equation}
\widetilde{F} \bigl( \eta \!-\! \eta' \big| k \bigr)
	=
	\frac{\cos \bigl[ k(\eta \!-\! \eta') \bigr]}{k}
	\biggl[
	\frac{1}{2}
	+
	\frac{1 }{ e^{k/T_{\rm rh}} - 1 }
	\biggr]
	\, ,
\end{equation}
where~$T_{\rm rh}$ is the reheating temperature.
We can thus write~\eqref{PhotonStatistical} in Fourier space,
\begin{equation}
\bigl\langle \bigl\{
	F_{\mu\nu}^{\scr (0)}(x) , F_{\rho\sigma}^{\scr (0)}(x')
	\bigr\} \bigr\rangle
	=
	\int \! \frac{d^3 \vec{k} }{(2 \pi)^3} \,
	e^{ i \vec{k} \cdot (\vec{x} - \vec{x}') }
	\bigl( 8 k_{[\mu} \eta_{\nu][\sigma} k_{\rho]} \bigr)
	\widetilde{F}
	\bigl( \eta \!-\! \eta' \big| k \bigr)
	\, ,
\end{equation}
where $k^\mu \!=\! (k^0, \vec{k}) \!=\! ( \|\vec{k}\| , \vec{k} )$.

The coincidence limit of the field strength correlator determines the 
expectation value of the energy-momentum tensor~\eqref{T0expectationFormal},
that sources the background expansion via~\eqref{BackgroundEinsteinEq}. 
In taking the coincidence limit we keep only the thermal contribution,
assuming the large temperature limit which we verify a posteriori to be consistent. 
In particular, we discard all ultraviolet divergences which we assume have been taken
care of by suitable counterterms that cannot depend on the temperature. This yields
\begin{equation}
\bigl\langle \bigl\{ F_{\mu\nu}^{\scr (0)}(x), F_{\rho\sigma}^{\scr (0)}(x) \bigr\} \bigr\rangle 
	\xrightarrow{ T_{\rm rh} \gg H_{\rm rh} }
	\! \int \! \frac{d^3 \vec{k} }{(2 \pi)^3} 
	\frac{ 8 k_{[\mu} \eta_{\nu][\sigma} k_{\rho]} }{k} 
	\frac{1}{e^{k/ T_{\rm rh} } - 1}
	=
	\Bigl(
	2 \delta_{[\mu}^0 \overline{\eta}_{\nu][\sigma} \delta_{\rho]}^0 
	+
	\overline{\eta}_{\mu[\rho} \overline{\eta}_{\sigma]\nu} \Bigr)
	\frac{4 \pi^2 T_{\rm rh}^4}{45}
	\, ,
\label{Fcorrelator}
\end{equation}
where~$\overline{\eta}_{\mu\nu} \!=\! \eta_{\mu\nu} \!+\! \delta_\mu^0 \delta_\nu^0$.
Then we can compute the expectation value of the background
energy-momentum tensor~\eqref{T0expectationFormal},
\begin{equation}
\label{T0expectationThermal}
\bigl\langle T_{\mu\nu}^{\scr (0)} \bigr\rangle 
	=
	\Bigl( 3 \delta_{\mu}^0 \delta_{\nu}^0 + \overline{\eta}_{\mu\nu} \Bigr) 
	\frac{\pi^2 T_{\rm rh}^4}{45 a^2} 
	=
	\Bigl( 4 u_\mu u_\nu + g_{\mu\nu} \Bigr) \frac{\pi^2 T_{\rm rh}^4}{45} \,,
\end{equation}
where in the second equality we have defined the normalized 
four-velocity~$u_\mu \!=\! - a \delta^0_\mu$. 
Combining this with the background Einstein tensor in~\eqref{zerothG} 
gives the background solution,
\begin{equation}
\epsilon = 2
\qquad \Longrightarrow \qquad 
\mathcal{H} = \frac{H_{\rm rh}}{a} 
\qquad \Longrightarrow \qquad 
H_{\rm rh}^2 = \frac{\pi^2 \kappa^2 T_{\rm rh}^4}{90} \, .
\label{THconnection}
\end{equation}
From here we see that the assumption $T_{\rm rh} \gg H_{\rm rh}$ is justified
as long as $(\kappa H_{\rm rh})^2 \ll 1$, which indeed is the
case for the radiation dominated universe, as explained in
Sec.~\ref{sec: Standard results for tensors in the early universe}.~\footnote{
The large temperature limit also allows us to neglect any contribution coming from the trace anomaly (e.g.~\cite{Brown:1976wc,Brown:1977pq}), whose contribution to the energy-momentum 
tensor~(\ref{T0expectationThermal}) is proportional to~$\eta_{\mu\nu} H_{\rm rh}^4 a^{-6}$.}

The coincidence limit of the field strength correlator also appears
in the Hartree term of the equation of motion~\eqref{HartreeRemovedEOM}
for the metric fluctuations. It is computed by appropriately contracting the 
tensor structure~\eqref{Wtensor} and the metric petrurbation 
into~\eqref{Fcorrelator},
\begin{equation}
- \frac{\kappa^2}{4a^2}
	{W_{\mu\nu}}^{\rho\sigma\alpha\beta\gamma\delta}
	\bigl\langle \bigl\{ F_{\rho\alpha}^{\scr (0)} ,
		F_{\sigma\beta}^{\scr (0)} \bigr\} \bigr\rangle
	h_{\gamma\delta}
	=
	-
	\mathcal{H}^2
	\Bigl(
	h_{\mu\nu}
	-
	\frac{1}{2} \eta_{\mu\nu} h
	-
	2 \delta_{\mu}^0 \delta_{\nu}^0 h
	-
	4 \delta_{(\mu}^0 h_{\nu)0}
	\Bigr)
	\, .
\label{HartreeTerm}
\end{equation}
where we have used the last two relations from~(\ref{THconnection}) to recognize 
that this contribution, corresponding to the 4-vertex diagram given in Fig.~\ref{4vertexDiag},
shows no~$\kappa^2$ suppression expected by naive power counting.
\begin{figure}[h!]
\centering
\includegraphics[width=4.3cm]{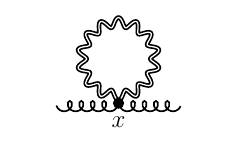}
\\
\vskip-4mm
\caption{1PI 4-vertex diagram contributing to the graviton equation of motion. 
The double wavy line stands for the statistical two-point function of the photon, 
while curly lines denote amputated gravitons. This diagram is conventionally defined as the
expectation value of the second variation of the Maxwell action with respect to
the metric, which does not correspond directly to the contribution in~(\ref{HartreeTerm}),
that is defined by perturbing the Einstein equation with the square root of the metric
stripped off. This detail is immaterial in practice 
as the same additional term not 
included~(\ref{HartreeTerm}) is also omitted in the left-hand side when perturbing
the Einstein tensor.}
\label{4vertexDiag}
\end{figure}
%

\subsection{Transverse traceless projection}

We are primarily interested in the propagation of transverse traceless 
perturbations, i.e.~tensor perturbations. We isolate these from the 
general equation~(\ref{HartreeRemovedEOM}) with the help of the transverse
traceless projector,
\begin{equation}
\Pi_{ijkl}
    =
    \Pi_{i(k} \Pi_{l)j} - \frac{1}{2} \Pi_{ij} \Pi_{kl}
    \, ,
\label{TTprojector}
\end{equation}
defined in terms of the idempotent transverse projector,
\begin{equation}
\Pi_{ij} = \delta_{ij}
    - \frac{\partial_i \partial_j }{\nabla^2} \, ,
\qquad
\Pi^{ij} \partial_j = \partial_j \Pi^{ij} = 0 \, ,
\qquad
\Pi_{ij} \Pi^{jk} = {\Pi_i}^k \, ,
\qquad
{\Pi_i}^i = 2 \, .
\end{equation}
The tensor perturbation is then defined as
\begin{equation}
\gamma_{ij}
    =
    {\Pi_{ij}}^{kl} h_{kl} \, .
\end{equation}
Acting with the projector~(\ref{TTprojector}) onto the equation of 
motion~\eqref{HartreeRemovedEOM} we get the equation of motion for
the tensor perturbation,
\begin{equation}
    \Bigl(
    \partial_0^2 +
    2 \mathcal{H} \partial_0 - \nabla^2
    \Bigr) \gamma_{ij}
    =
    \frac{1}{a^2}
    {\Pi_{ij}}^{kl} S_{kl}
    \, .
\label{TTeq}
\end{equation}
We should note here that the left-hand side of the equation reveals
the graviton to be massless. This is due to the Hartree 
term~(\ref{HartreeTerm}), depicted in Fig.~\ref{4vertexDiag},
canceling the would-be mass term from the 
perturbation of the Einstein tensor~(\ref{EinsteinPerturbation}).

We aim to solve Eq.~(\ref{TTeq}) treating the source as a perturbation.
That is facilitated by introducing another retarded scalar Green's function,
\begin{equation}
\Bigl(
    \partial_0^2 +
    2 \mathcal{H} \partial_0 - \nabla^2 
    \Bigr)
    G_{\scr \rm R}^{\scr \rm TT}(x;x')
    =
    \frac{\delta^4(x \!-\! x')}{a^2} \, ,
\label{GTTeq}
\end{equation}
that had been worked out in~\cite{Ota:2023iyh}. Here we choose
the scale factor normalization on the right-hand side for 
convenience, so that the Green's function absorbs all the explicit powers 
of the scale factors in loop computations, and so that the Green's function
is symmetric in time arguments, apart from the overall step function.
The solution for this Green's function is best expressed in 
spatial momentum space,
\begin{equation}
G_{\scr \rm R}^{\scr \rm TT}(x;x')
	=
	\int\! \frac{d^3 \vec{k} }{(2\pi)^3} \, 
	e^{i \vec{k} \cdot (\vec{x} - \vec{x}')} \,
	\widetilde{G}_{\scr \rm R}^{\scr \rm TT}
	\bigl( \eta; \eta' \big| k \bigr)
	\, ,
\end{equation}
where the Green's function reads
\begin{equation}
\widetilde{G}_{\scr \rm R}^{\scr \rm TT}
    \bigl( \eta; \eta' \big| k \bigr)
    =
    \frac{ \theta(\eta \!-\! \eta') }{a(\eta)a(\eta')}
    \frac{ \sin\bigl[ k(\eta \!-\! \eta') \bigr] }{ k }
    \, .
\end{equation}
This allows us to cast Eq.~(\ref{TTeq}) in the form of the Yang-Feldman
equation,
\begin{equation}
\gamma_{ij}(x)
    =
    \gamma_{ij}^{0}(x)
    +
    \int\! d^4x' \,
    {G}_{\scr \rm R}^{\scr \rm TT} (x;x')
    {\Pi_{ij}}^{kl}(\vec{x}')
    S_{kl}(x')
    \, ,
\label{YFeq}
\end{equation}
where~$\gamma_{ij}^0$ is the tree-level solution. This is now an integral 
equation that is well adapted to a perturbative expansion, that is obtained 
by iterating the equation.

\subsection{Tensor perturbation correlator}

We want to compute the one-loop correction to the statistical two-point 
functions of tensor perturbations,
\begin{equation}
\bigl\langle \bigl\{ \gamma_{ij}(x) , \gamma_{kl}(x') \bigr\} \bigr\rangle
	=
	\int\! \frac{ d^3 \vec{k} }{ (2\pi)^3 } \,
	e^{i \vec{k} \cdot ( \vec{x} - \vec{x}' )}
	{\mathbb{P}}_{ijkl}(\vec{k}) \times
	P\bigl( \eta; \eta' \big| k \bigr)
	\, ,
\label{h2pt}
\end{equation}
where the momentum space transverse traceless projector is
\begin{equation}
\mathbb{P}_{ijkl}(\vec{k})
	=
	\mathbb{P}_{i(k}(\vec{k}) \mathbb{P}_{l)j}(\vec{k})
	-
	\frac{1}{2} \mathbb{P}_{ij}(\vec{k}) \mathbb{P}_{kl}(\vec{k})
	\, ,
\qquad \quad
\mathbb{P}_{ij}(\vec{k})
	=
	\delta_{ij} - \frac{\vec{k}_i \vec{k}_j}{\vec{k}^2}
	\, ,
\end{equation}
In particular, we are interested in the dimensionless counterpart of the
momentum space two-point function, expressed in terms of the dimensionful
one introduced in~(\ref{h2pt}) as
\begin{equation}
\mathcal{P} \bigl( \eta;\eta' \big| k \bigr)
    = \frac{\kappa^2 k^3}{2\pi^2} P \bigl( \eta;\eta' \big| k \bigr) \, ,
\label{dimensionless2pt}
\end{equation}
whose coincident limit reduces to the definition of the dimensionless
tensor power spectrum defined in Sec.~\ref{sec: Standard results for tensors in the early universe},
$\mathcal{P} \bigl( \eta;\eta \big| k \bigr) \!=\! \mathcal{P}_t(\eta,k)$.

By plugging in the Yang-Feldman equation~\eqref{YFeq} into the definition~(\ref{h2pt}),
and applying Wick's theorem, we can collect all the one-loop contributions,
\begin{align}
\MoveEqLeft[3]
\bigl\langle \bigl\{ \gamma_{ij}(x) , \gamma_{kl}(x') \bigr\} \bigr\rangle
	=
	\bigl\langle \bigl\{ \gamma_{ij}^0(x) , \gamma_{kl}^0(x') \bigr\} \bigr\rangle
\label{contribution0}
\\
&
	+
	\int\! d^4y \,
	G_{\scr \rm R}^{\scr \rm TT} (x;y)
	\int\! d^4y' \,
	G_{\scr \rm R}^{\scr \rm TT} (x';y')
	{\Pi_{ij}}^{mn}(\vec{y})
	{\Pi_{kl}}^{ab}(\vec{y}')
	\bigl\langle \bigl\{ S_{mn}^{\scr \rm I}(y) , 
		S_{ab}^{\scr \rm I}(y') \bigr\} \bigr\rangle
\label{contributionA}
\\
&
	+
	\int\! d^4y \,
	G_{\scr \rm R}^{\scr \rm TT} (x;y)
	{\Pi_{ij}}^{mn}(\vec{y})
	\bigl\langle \bigl\{ S_{mn}^{\scr \rm III}(y) ,
		\gamma_{kl}^0(x') \bigr\} \bigr\rangle
\label{contributionB}
\\
&
	+
	\int\! d^4y' \,
	G_{\scr \rm R}^{\scr \rm TT} (x';y')
	{\Pi_{kl}}^{ab}(\vec{y}')
	\bigl\langle \bigl\{ \gamma_{ij}^0(x) , S_{ab}^{\scr \rm III}(y')
	\bigr\} \bigr\rangle
	\, ,
\label{contributionC}
\end{align}
where the correlators involving sources~$S_{ij}^{\scr \rm I-III}$ are given by:
\begin{align}
\bigl\langle \bigl\{ S_{mn}^{\scr \rm I}(y) , S_{ab}^{\scr \rm I}(y') \bigr\} \bigr\rangle
	={}&
	\kappa^2 ( a_y a_{y'})^2
	\Bigl\langle \Bigl\{
	T_{mn}^{\scr (0)}(y) 
		- \bigl\langle T_{mn}^{\scr (0)}(y) \bigr\rangle  ,
	T_{ab}^{\scr (0)}(y') 
		- \bigl\langle T_{ab}^{\scr (0)}(y') \bigr\rangle
	\Bigr\} \Bigr\rangle
\nonumber \\
={}&
	\kappa^2
	\Bigl( \delta_{(m}^\mu \delta_{n)}^\nu
		- \frac{1}{4} \eta_{mn} \eta^{\mu\nu} \Bigr) \eta^{\alpha\beta}
	\Bigl( \delta_{(a}^\rho \delta_{b)}^\sigma
		- \frac{1}{4} \eta_{ab} \eta^{\rho\sigma} \Bigr) \eta^{\omega\lambda}
\nonumber \\
&   \hspace{.5cm}
	\times
	\Bigl[
	\bigl\langle \bigl\{ 
		F_{\mu\alpha}^{\scr(0)}(y) , F_{\rho\omega}^{\scr(0)}(y')
		\bigr\} \bigr\rangle
	\bigl\langle \bigl\{ 
		F_{\nu\beta}^{\scr (0)}(y) , F_{\sigma\lambda}^{\scr (0)}(y')
		\bigr\} \bigr\rangle
\nonumber \\
&   \hspace{1.5cm}
	+
	\bigl\langle \bigl[
		F_{\mu\alpha}^{\scr(0)}(y) , F_{\rho\omega}^{\scr(0)}(y')
		\bigr] \bigr\rangle
	\bigl\langle \bigl[
		F_{\nu\beta}^{\scr (0)}(y) , F_{\sigma\lambda}^{\scr (0)}(y')
		\bigr] \bigr\rangle
	\Bigr]
	\, ,
\label{SISI}
\\
\bigl\langle \bigl\{ S_{mn}^{\scr \rm III}(y) , \gamma_{kl}^0(x') \bigr\} \bigr\rangle
	={}&
	\kappa^2
	\Bigl( \delta_{(m}^\mu \delta_{n)}^\nu
		- \frac{1}{4} \eta_{mn} \eta^{\mu\nu}
		\Bigr) \eta^{\alpha\beta} 
	\int\! d^4y' \, G_{\scr \rm R}(y - y')
\nonumber
\\
&
	\times 
	\partial^{y'}_\omega
	\partial^{y'}_{[\nu}
	\Bigl[ {V_{\beta]}}^{\rho\omega\sigma ab} 
		\bigl\langle \bigl\{ F_{\mu\alpha}^{\scr (0)}(y) , 
			F_{\rho\sigma}^{\scr (0)}(y') \bigr\} \bigr\rangle
	\bigl\langle \bigl\{ \gamma_{ab}^0(y') , 
		\gamma_{kl}^0(x') \bigr\} \bigr\rangle
		\Bigr]
    \, ,
\\
\bigl\langle \bigl\{ \gamma_{ij}^0(x) , S_{ab}^{\scr \rm III}(y') \bigr\} \bigr\rangle
	={}&
	\kappa^2
	\Bigl( \delta_{(a}^\rho \delta_{b)}^\sigma
		- \frac{1}{4} \eta_{ab} \eta^{\rho\sigma}
		\Bigr) \eta^{\omega\lambda} 
	\int\! d^4y \, G_{\scr\rm R}(y' - y)
\nonumber
\\
&
	\times
	\partial^{y}_\alpha
	\partial^{y}_{[\sigma}
	\Bigl[
	{V_{\lambda]}}^{\mu\alpha\nu mn} 
	\bigl\langle \bigl\{ F_{\rho\omega}^{\scr (0)}(y') , 
		F_{\mu\nu}^{\scr (0)}(y) \bigr\} \bigr\rangle
	\bigl\langle \bigl\{ \gamma_{ij}^0(x) , 
		\gamma_{mn}^0(y) \bigr\} \bigr\rangle \Bigr]
	\, .
\end{align}
Note that of the three contributions $S_{mn}^{\scr \rm I}$, $S_{mn}^{\scr \rm II}$ 
and $S_{mn}^{\scr \rm III}$ only the first and the last contribute to order $\kappa^2$. 
Namely, terms involving two of $S_{mn}^{\scr \rm II}$ and/or $S_{mn}^{\scr \rm III}$ 
are of order $\kappa^4$, while the mixed 
terms~$\bigl\langle \bigl\{ S_{mn}^{\scr \rm II}(y),\gamma_{kl}^0(x') \bigr\} \bigr\rangle = 0$ vanish.

The contribution in~\eqref{contribution0} is just the tree-level
two-point function. The three contributions of order~$\kappa^2$
in~\eqref{contributionA}--\eqref{contributionC} are one-loop contributions 
given diagrammatically in Fig.~\ref{diagrams}, and labeled respectively
as~$A$, $B$, and $C$.
Contribution~$A$ accounts for the so-called thermally induced gravitational 
wave production~\cite{Ghiglieri:2015nfa,Ghiglieri:2020mhm}. This 
contribution does not depend on primordial tensor perturbations, 
and is there regardless of them, purely because of the thermal nature of the 
photon thermal medium. Contributions~$B$ and~$C$ depend linearly on the
primordial tensor perturbation and represent the modulation of the
primordial signal. In the following section we compute these three 
one-loop diagrams.

\begin{figure}[h!]
\centering
\hfill
\includegraphics[width=4.9cm]{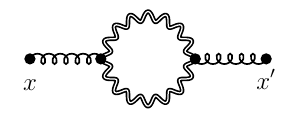}
\hfill
\includegraphics[width=4.9cm]{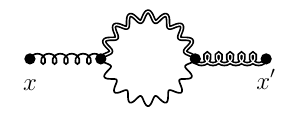}
\hfill
\includegraphics[width=4.9cm]{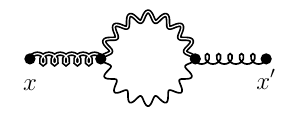}
\hfill \
\vskip-1mm
\hfill $A$ \hfill\hfill $B$ \hfill\hfill \ $C$ \hfill \,\!
\\
\vskip-2mm
\caption{One-loop corrections to the graviton two-point function.
Curly represent gravitons, while wavy lines represent photons.
Double lines stand for statistical two-point functions, while single
lines stand for retarded propagators.
Diagram $A$ depicts the so-called thermally induced gravitational
wave production. Diagrams $B$ and $C$ depict contributions dubbed
radiation exchange in~\cite{Ota:2023iyh}.}
\label{diagrams}
\end{figure}
%

\section{Computing diagrams}
\label{sec: Computing diagrams}

This section is devoted to computing the three diagrams in 
Fig.~\ref{diagrams}. They are computed in momentum space, in the 
superhorizon limits, and in the late time limit. For a more detailed 
computation and a discussion of various limits see~\cite{Frob:2025uev}.

\subsection{Thermal GW production}

Type~$A$ contribution in~(\ref{contributionA}) written out explicitly
reads
\begin{align}
\bigl\langle \bigl\{ \gamma_{ij}(x) , \gamma_{kl}(x') \bigr\} \bigr\rangle^A
	={}&
	\kappa^2
	\int\! d^4y \,
	{G}_{\scr \rm R}^{\scr \rm TT} (x;y)
	\int\! d^4y' \,
	{G}_{\scr \rm R}^{\scr \rm TT} (x';y') \,
	{\Pi_{ij}}^{mn}(\vec{y})
	{\Pi_{kl}}^{ab}(\vec{y}')
\nonumber \\
&
	\times
	\eta^{\mu\nu} \eta^{\rho\sigma}
	\bigl\langle \bigl\{ 
	F_{m\mu}^{\scr(0)}(y) , F_{a\rho}^{\scr(0)}(y')\bigr\} \bigr\rangle
	\bigl\langle \bigl\{ F_{n\nu}^{\scr (0)}(y) , F_{b\sigma}^{\scr (0)}(y')
	\bigr\} \bigr\rangle
	\, .
\end{align}
where we have discarded the term in the last line of~(\ref{SISI})
on account of it not containing any temperature dependence.
Expressed in momentum space, in terms of the dimensionless two-point 
function~(\ref{dimensionless2pt}), this contribution reads,
\begin{align}
\MoveEqLeft[3]
\Delta \mathcal{P}^{A} \bigl( \eta;\eta' \big| k \bigr)
	=
    \frac{2\kappa^4 k^3}{\pi^2} \!
	\int_{\eta_{\rm rh}}^{\eta}\! d\eta_y \,
	\widetilde{G}_{\scr \rm R}^{\scr \rm TT} \bigl( \eta; \eta_y \big| k \bigr)
	\int_{\eta_{\rm rh}}^{\eta'}\! d\eta_{y'} \,
	\widetilde{G}_{\scr \rm R}^{\scr \rm TT}
	\bigl( \eta'; \eta_{y'} \big| k \bigr)
\nonumber \\
&
	\times\!
	\int \! \frac{d^3 \vec{q}}{(2 \pi)^3} \,
	\widetilde{F} \bigl( \eta_y \!-\! \eta_{y'} \big| q \bigr)
	\widetilde{F} \bigl( \eta_y \!-\! \eta_{y'} \big| \| \vec{k} \!-\! \vec{q} \| \bigr)
	\times
	\mathbb{P}_{mnab}(\vec{k})
\nonumber \\
&   \hspace{1.cm}
	\times\! 
	\Bigl[
	q^{m} q^{n} q^{a} q^{b}
	+
	2 q^{(m} \eta^{n)(a} q^{b)} (k \!-\! q)^\mu q_\mu
	+
	\frac{1}{2} (k \!-\! q)^\mu q_\mu (k \!-\! q)^\nu q_\nu \eta^{m(a} \eta^{b)n}
	\Bigr]
	\, .
\end{align}
where $(k-q)^\mu q_\mu = \|\vec{q}\| \|\vec{k}-\vec{q}\| + \vec{q} \cdot ( \vec{k} - \vec{q} )$.
We are interested in the small momentum limit~$k \!\ll\! \mathcal{H}$,
corresponding to superhorizon modes, for which the expression reduces to
\begin{align}
\Delta \mathcal{P}^{A} \bigl( \eta;\eta' \big| k \bigr)
	\ \overset{k \ll \mathcal{H}}{\longsim}& \
    \frac{2\kappa^4 k^3}{\pi^2} \!
	\int_{\eta_{\rm rh}}^{\eta}\! d\eta_y \,
	\widetilde{G}_{\scr \rm R}^{\scr \rm TT} \bigl( \eta; \eta_y \big| 0 \bigr)
	\int_{\eta_{\rm rh}}^{\eta'}\! d\eta_{y'} \,
	\widetilde{G}_{\scr \rm R}^{\scr \rm TT}
	\bigl( \eta'; \eta_{y'} \big| 0 \bigr)
\nonumber \\
&
	\times \!
	\int \! \frac{d^3 \vec{q} }{(2 \pi)^3} \,
	\Bigl[ \widetilde{F} \bigl( \eta_y \!-\! \eta_{y'} \big| q \bigr) \Bigr]^2
	\times
	\Bigl[
	\mathbb{P}_{mnab}(\vec{k}) q^{m} q^{n} q^{a} q^{b}
	\Bigr]_{\vec{k} \to 0}
	\, .
\end{align}
This limit is assumed in the remainder of this subsection.
The loop integral over~$\vec{q}$ is computed in the Appendix~\ref{appendix:loopintegral}, which leaves the two integrals over
retarded Green's functions to evaluate. However, since we are not interested
in the details of the transient contributions that quickly become irrelevant,
but only in the leading contributions, it is more convenient to write
the problem as a double differential equation
making use of the equation of motion~(\ref{GTTeq}),
\begin{align}
\Bigl( \partial_0^2 + 2 \mathcal{H} \partial_0 \Bigr)
    \Bigl( \partial_0'^2 + 2 \mathcal{H}' \partial_0' \Bigr)
    \Delta \mathcal{P}^{A} \bigl( \eta ; \eta' \big| k \bigr)
    ={}&
    \frac{2\kappa^4 k^3}{\pi^2 (aa')^2} 
        I_1 \bigl( T_{\rm rh},\Delta\eta \bigr)
        \, ,
\end{align}
where the~$I_1$ is given in Eq.~(\ref{I1solution}), and 
where~$\Delta\eta\!=\!\eta\!-\!\eta'$. Ultimately we are interested in the
time coincidence limit of the tensor two-point function, for which it
is sufficient to consider the equation above close to time coincidence
(adopting the strategy from~\cite{Glavan:2019uni}),
\begin{align}
\Bigl( \partial_0^2 + 2 \mathcal{H} \partial_0 \Bigr)
    \Bigl( \partial_0'^2 + 2 \mathcal{H}' \partial_0' \Bigr)
    \Delta \mathcal{P}^A \bigl( \eta ; \eta' \big| k\bigr)
    ={}&
    \frac{32 \kappa^2 k^3 T_{\rm rh} H_{\rm rh}^2  }{5\pi^2 (aa')^2 }
        \Bigl[
        1 + \mathcal{O} \bigl( \Delta\eta^2 \bigr)
        \Bigr]
        \, .
\end{align}
The solution at late times is then simply found to be
\begin{equation}
\Delta \mathcal{P}^A \bigl( \eta ; \eta' \big| k \bigr)
    =
    \frac{32 \kappa^2 k^3 T_{\rm rh}}{5 \pi^2 H_{\rm rh}^2}
    \times
    \ln\Bigl( \frac{a}{a_{\rm rh}} \Bigr) \ln\Bigl( \frac{a'}{a_{\rm rh}} \Bigr)
    \, ,
\end{equation}
while the coincidence limit gives the desired correction to the
superhorizon tensor power spectrum,
\begin{equation}
\frac{\Delta \mathcal{P}_t^A }{\mathcal{P}_t^0}
    =
    \frac{32 }{5}
    \times
    \Bigl( \frac{k^3}{H_{\rm rh}^3} \Bigr)
    \times
    \Bigl( \frac{T_{\rm rh}}{H_\text{inf}} \Bigr)
    \times
    \Bigl( \frac{H_{\rm rh}}{H_\text{inf}} \Bigr)
    \times
    \ln^2\Bigl( \frac{a}{a_{\rm rh}} \Bigr)
    \, .
\end{equation}
While we obtain a different result for this contribution compared 
to the respective result in~\cite{Ota:2023iyh}, this is of little consequence
as this contribution is negligible for observable scales
compared to the tree-level spectrum 
(cf. the discussion in
Sec.~\ref{sec: Standard results for tensors in the early universe}),
on account of the~$(k/H_{\rm inf})^3$ suppression,
despite the enhancement due to~$T_{\rm rh}/H_{\rm inf}$.

\subsection{Radiation exchange}
\label{subsec: Radiation exchange}

Plugging in the tensor structure~\eqref{Vtensor} into contribution~$B$
in~\eqref{contributionB}, using Maxwell's equation, and partially 
integrating some spatial derivatives gives
\begin{align}
\MoveEqLeft[4]
\bigl\langle \bigl\{ \gamma_{ij}(x) , \gamma_{kl}(x') \bigr\} \bigr\rangle^B
    =
    \kappa^2 \!
    \int\! d^4y \,
    G_{\scr \rm R}^{\scr \rm TT}(x;y)
    {\Pi_{ij}}^{mn}(\vec{y})
    \int\! d^4y' \,
\nonumber
\\
&
    \times \!
    \,
    \Bigl[
    \partial^{y'}_{n} \partial_{y'}^a
    G_{\scr \rm R}(y \!-\! y')
    \times
        \bigl\langle \bigl\{ 
        F_{m\omega}(y) , F^{b \omega}(y') 
        \bigr\} \bigr\rangle^{\scr (0)}
    \times
    \bigl\langle \bigl\{ \gamma_{ab}^0(y') , 
        \gamma_{kl}^0(x') \bigr\} \bigr\rangle
\nonumber \\
&   \hspace{0.8cm}
    +
    \partial^{y'}_{n}
    G_{\scr \rm R}(y \!-\! y')
    \times
    \bigl\langle \bigl\{ {F_{m}}^a(y) , F^{b \omega}(y') 
        \bigr\} \bigr\rangle^{\scr (0)}
    \times
    \partial^{y'}_\omega
    \bigl\langle \bigl\{ \gamma_{ab}^0(y') , 
        \gamma_{kl}^0(x') \bigr\} \bigr\rangle
\nonumber \\
&   \hspace{0.8cm}
    +
    \partial_{y'}^a
    G_{\scr \rm R}(y \!-\! y')
    \times
        \bigl\langle \bigl\{ {F_{m}}^\omega(y) , {F^b}_n(y') 
        \bigr\} \bigr\rangle^{\scr (0)}
    \times
    \partial^{y'}_{\omega} 
    \bigl\langle \bigl\{ \gamma_{ab}^0(y') , 
        \gamma_{kl}^0(x') \bigr\} \bigr\rangle
\nonumber \\
&   \hspace{0.8cm}
    +
    G_{\scr \rm R}(y \!-\! y')
    \times
    \bigl\langle \bigl\{ {F_{m}}^\omega(y) , F^{b \lambda}(y') 
        \bigr\} \bigr\rangle^{\scr (0)}
    \times
    \partial^{y'}_\omega \partial^{y'}_{\lambda}
    \bigl\langle \bigl\{ \gamma_{nb}^0(y') , 
        \gamma_{kl}^0(x') \bigr\} \bigr\rangle
    \Bigr]
    \, .
\end{align}
Taking the spatial Fourier transform then gives
\begin{align}
\MoveEqLeft[3]
\Delta\mathcal{P}^B \bigl( \eta; \eta' \big| k \bigr)
    =
    -
    \kappa^2 \!
    \int_{\eta_{\rm rh}}^{\eta}\! d\eta_y \,
	\widetilde{G}_{\scr \rm R}^{\scr \rm TT}\bigl( \eta; \eta_y \big| k \bigr)
    \int_{\eta_{\rm rh}}^{\eta_y}\! d\eta_{y'}
     \int\! \frac{d^3\vec{q} }{(2\pi)^3} \, 
        \widetilde{G}_{\scr \rm R}\bigl( \eta_y \!-\! \eta_{y'} \big| q \bigr)
\nonumber \\
&
\times\!
\biggl\{
\widetilde{F}\bigl( \eta_y \!-\! \eta_{y'} \big| \| \vec{k} \!-\! \vec{q} \| \bigr)
    \biggl[
    2 q^i q^j q^k q^l \mathbb{P}_{ijkl}(\vec{k}) 
	+
    q^i q^j \mathbb{P}_{ij}(\vec{k}) 
    \bigl[ ( \partial^{y'}_0 )^2 + k^2 \bigr]
\nonumber \\
&   \hspace{4.5cm}
	+
    (k\!-\!q)^i k_i (k\!-\!q)^j k_j
    +
    2 \| \vec{k} \!-\! \vec{q} \|^2
    ( \partial^{y'}_0 )^2
	\biggr]
    \mathcal{P} \bigl( \eta_{y'}; \eta' \big| k \bigr)
\nonumber \\
&   \hspace{1cm}
	-
	\partial_{0}^y  \widetilde{F}\bigl( \eta_y \!-\! \eta_{y'} \big| \| \vec{k} \!-\! \vec{q} \| \bigr)
    \times
    2 q^i q^j {\mathbb{P}}_{ij}(\vec{k})
    \times
    \partial^{y'}_0
	\mathcal{P}^0 \bigl( \eta_{y'}; \eta' \big| k \bigr)
\biggr\}
    \, ,
\end{align}
where~$\mathcal{P}^0$ is the tree-level dimensionless two-point function.
Finally, we are interested in the superhorizon 
limit~$k \!\ll\!\mathcal{H}$, henceforth assumed until the end of the
subsection, in which the expression reduces to,
\begin{align}
\Delta \mathcal{P}^B\bigl( \eta; \eta' \big| k \bigr)
	\ \overset{k\ll \mathcal{H}}{\longsim}{}&
	- 2 \kappa^2 \!
	\int_{\eta_{\rm rh}}^{\eta}\! d\eta_{y} \,
	\widetilde{G}_{\scr \rm R}^{\scr \rm TT} \bigl( \eta; \eta_y \big| k \bigr)
    \mathcal{P}_0\bigl( \eta_y; \eta' \big| 0 \bigr)
\nonumber \\
&	\times\!
	\int_{\eta_{\rm rh}}^{\eta_y}\! d\eta_{y'} \!
	\int\! \frac{d^3 \vec{q} }{(2\pi)^3} \,
	\widetilde{G}_{\scr \rm R} \bigl( \eta_y \!-\! \eta_{y'} \big| q \bigr)
	\widetilde{F} \bigl( \eta_y \!-\! \eta_{y'} \big| q \bigr)
	\Bigl[ \mathbb{P}_{ijkl}(\vec{k}) q^i q^j q^k q^l \Bigr]_{\vec{k}\to0}
	\, .
\end{align}
where we used the fact that the kernel of the integral over~$\eta_{y'}$
has effectively a narrow support in order to extract the tree-level 
power spectum outside of that integral.
The loop integral over~$\vec{q}$ is the integral~$I_2$ evaluated in 
the Appendix in Eq.~(\ref{I2solution}). 
Instead of evaluating explicitly the remaining integral over the 
retarded Green's function, it is again advantageous to rewrite the
expression as a differential equation,
\begin{equation}
\Delta \mathcal{P}^B\bigl( \eta ; \eta' \big| 0 \bigr)
	=
	\mathcal{P}_0\bigl( \eta ; \eta' \big| 0 \bigr)
	\times
	\int_{\eta_{\rm rh}}^{\eta}\! d\eta_y \,
	\widetilde{G}_{\scr \rm R}^{\scr \rm TT} \bigl( \eta; \eta_y \big| k \bigr)
	\!\times\!
	\frac{ \kappa^2 }{120 \pi^2}
	\biggl[
	\Bigl( \frac{\partial}{\partial \Delta\eta} \Bigr)^{\!3}
	\mathcal{I}(T,\Delta\eta)
	\biggr]_0^{\eta_y - \eta_{\rm rh} }
	\, ,
\end{equation}
where we have evaluated the temporal integral over~$I_2$ using its
derivative representation~\eqref{TotalDerIntegral},
that close to coincidence reads
\begin{align}
\Bigl( \partial_0^2 + 2 \mathcal{H} \partial_0 \Bigr)
	\Delta \mathcal{P}^B\bigl( \eta ; \eta' \big| k \bigr)
	={}&
    \biggl[
	\frac{ \pi^2 \kappa^2 T_{\rm rh}^4 }{ 225 a^2 }
	+
	\mathcal{O} \bigl( \Delta\eta^2 \bigr)
    \biggr]
	\times
	\mathcal{P}_0\bigl( \eta ; \eta \big| 0 \bigr)
	=
	\frac{ 2 H_0^2 }{ 5 a^2 }
	\times
	\mathcal{P}_0\bigl( \eta ; \eta \big| 0 \bigr)
	\, .
\end{align}
Inverting this equation at late times is straightforward,
\begin{equation}
\Delta \mathcal{P}^B\bigl( \eta ; \eta' \big| 0 \bigr)
	=
	\frac{ 2 }{ 5 } \ln \Bigl( \frac{a}{a_{\rm rh}} \Bigr)
	\times 
	\mathcal{P}_0\bigl( \eta ; \eta' \big| 0 \bigr)
	\, .
\end{equation}
and the~$C$ type contribution is obtained by simply exchanging time arguments,
\begin{equation}
\Delta \mathcal{P}^{C} \bigl( \eta ; \eta' \big| 0 \bigr)
	=
	\frac{ 2 }{ 5 } \ln \Bigl( \frac{a'}{a_{\rm rh}} \Bigr)
	\times 
	\mathcal{P}_0\bigl( \eta ; \eta \big| 0 \bigr)
	\, .
\end{equation}
Adding the two contributions together and taking the coincidence limit
gives the full correction to tensor power spectrum,
\begin{equation}
\frac{ \Delta \mathcal{P}_t^{B+C} }{ \mathcal{P}_t^0 }
    = \frac{4}{5} 
    \ln \Bigl( \frac{a}{a_{\rm rh}} \Bigr)
    \, .
\label{DeltaPtB+C}
\end{equation}
This result matches that reported in~\cite{Ota:2023iyh}. The most striking feature of this 
result is the absence of any suppression
compared to the tree-level result, which signals the breakdown of 
perturbation theory, and points to potentially interesting physics,
provided that a reliable way of quantifying the correction is found.

\section{Resummed computation}
\label{sec: Resummed computation}

The perturbative results for the corrections found in the preceding section
show two features for~$B$ and~$C$ type corrections:
(i) they are not suppressed by a small parameter, and (ii) they grow in time.
This points to the fact that treating these corrections perturbatively
is not justified. Rather, these corrections are parametrically of the same
order as the terms in the part of the linear equation that determines the tree-level
solution. That is why here we explore the limits of the linear approximation
for the evolution of tensor modes, by treating~$B$ and~$C$ type corrections 
on the same footing. This is facilitated by the~$A$ type contribution
being negligible, which allows to write a homogeneous linear equation for 
the tensor two-point function,
\begin{align}
&
    \Bigl(
    \partial_0^2 +
    2 \mathcal{H} \partial_0 - \nabla^2
    \Bigr) \bigl\langle \bigl\{ \gamma_{ij}(x) 
        \gamma_{kl}(x') \bigr\} \bigr\rangle
    -
    \frac{ 2 \kappa^2 }{a^2}
    {\Pi_{ij}}^{mn}
	\eta^{\alpha\beta} 
\nonumber
\\
&   \hspace{0.5cm}
    \times \!
    \int\! d^4x'' \, G_{\scr \rm R}(x \!-\! x'')  \,
    \partial''_\kappa \partial''_{[m}
    \Bigl[ 
        {V_{\beta]}}^{\theta\kappa\lambda rs}
        \bigl\langle \bigl\{ F_{n\alpha}^{\scr (0)}(x) , 
            F_{\theta\lambda}^{\scr (0)} (x'') \bigr\} \bigr\rangle
        \bigl\langle \bigl\{ \gamma_{rs}(x'') 
            \gamma_{kl}(x') \bigr\} \bigr\rangle \Bigr]
    = 0 \, .
\end{align}
By writing this equation in momentum space, and applying the same 
approximations for the superhorizon limit ($k\!\ll\! \mathcal{H}$)
made in Sec.~\ref{subsec: Radiation exchange} we find the equation to
reduce to a local one,
\begin{equation}
\Bigl(
    \partial_0^2 +
    2 \mathcal{H} \partial_0
    +
    \frac{m_{\rm eff}^2}{a^2}
    \Bigr) 
    \mathcal{P}\bigl( \eta; \eta' \big| k \bigr)
    =
    0
    \, ,
\label{MassiveEq}
\end{equation}
where the original nonlocality is captured by the effective mass term,
\begin{align}
m_{\rm eff}^2
    ={}&
    \frac{2 \pi^3 \kappa^2 T_{\rm rh}^5}{15} 
    \int^{\eta-\eta_{\rm rh}}_0 \! d\Delta\eta \,
    \biggl[
        \frac{3}{8( \pi T_{\rm rh} \Delta\eta )^5}
        -
        \frac{11 \, {\rm ch} ( 2\pi T_{\rm rh} \Delta\eta ) 
            + {\rm ch} (6\pi T_{\rm rh} \Delta\eta) }{ {\rm sh}^5 (2\pi T_{\rm rh} \Delta\eta) }
        \biggr]
\nonumber \\
={}&
    \frac{\pi^2 \kappa^2 T_{\rm rh}^4}{15} 
    \biggl[
        \frac{2}{{\rm sh}^2(2\pi T_{\rm rh} \Delta\eta_{\rm rh})}
        +
        \frac{3}{{\rm sh}^4(2\pi T_{\rm rh} \Delta\eta_{\rm rh})}
        -
        \frac{3}{16( \pi T_{\rm rh} \Delta\eta_{\rm rh} )^4}
        -
        \frac{1}{15}
        \biggr]
        \, ,
\end{align}
with~$\Delta\eta_{\rm rh} \!=\! \eta \!-\! \eta_{\rm rh}$. 
We should emphasize that this mass term is not a local mass term
that is canceled by the Hartree approximation~\cite{Liu:2024utl},
but rather an approximation for the nonlocal correction to the 
linear equation of motion for the tensor perturbation.
Most of the terms in this effective mass term decay quickly, so that the 
late time approximation is warranted,
\begin{align}
m_{\rm eff}^2
    \xrightarrow{T_{\rm rh} \Delta\eta_{\rm rh} \gg 1}
    -
    \frac{\pi^2 \kappa^2 T_{\rm rh}^4}{225} 
    =
    -
    \frac{ 2 }{5} H_{\rm rh}^2
        \, .
\end{align}
In general the effective mass squared is proportional to 
the 4-dimensional momentum space retarded self-energy 
in momentum space.\footnote{The Fourier transforms need to
be performed in the correct sequence: first the spatial one, and then the 
temporal one. In general the two transforms are known not to commute in
thermal field theory~\cite{Rebhan:1990yr}.}
Finally, the simple equation governing the evolution of the 
tensor perturbation two-point function on superhorizon scales reads
\begin{equation}
\Bigl(
    \partial_0^2 +
    2 \mathcal{H} \partial_0
    -
    \lambda \mathcal{H}^2
    \Bigr) 
    \mathcal{P} \bigl( \eta; \eta' \big| k \bigr)
    =
    0
    \, ,
\qquad\quad
\lambda = \frac{2}{5} \, .
\end{equation}
It is evident from this form of the equation that treating parts
of it perturbatively is not warranted, as there is no naive~$\kappa^2$ 
suppression that remains. The late time solution is found to be
(remembering that another equation with respect to the primed coordinate must
also hold),
\begin{equation}
\mathcal{P}\bigl( \eta;\eta' \big| k \bigr)
    =
    \mathcal{P}_0 \bigl( \eta;\eta' \big| k \bigr)
    \!\times\!
    \Bigl( \frac{aa'}{a_{\rm rh}^2} \Bigr)^{ \! \frac{- 1 + \sqrt{ 1 + 4\lambda } }{2} }
    \, ,
\end{equation}
where the tree-level two-point function is essentially time-independent.
This means that the growth of the tensor power spectrum turns into a 
power-law~\footnote{This result seems to have been confirmed numerically
in~\cite{Ota:2025yeu}.}
\begin{equation}
\mathcal{P}_t
    =
    \mathcal{P}_t^0
    \times
    \Bigl( \frac{a}{a_{\rm rh}} \Bigr)^{ \!- 1 + \sqrt{ 1 + 4\lambda } }
    \, ,
\label{PowerLawCorrection}
\end{equation}
with the exponent is~$(- 1 \!+\! \sqrt{ 1 + 4\lambda })\!\approx\! 0.612$.
This correctly reproduces the perturbative result 
in~(\ref{DeltaPtB+C}) when formally expanded for small~$\lambda$ to
linear order, but the resummation shows a much faster growth than 
the logarithmic one obtained by the one-loop approximation.

\section{Discussion}
\label{sec: Discussion}

Interpreting the fluctuations of the cosmic microwave background as
having their origin in primordial inflation hinges on the assumption
of conservation of primordial fluctuations on superhorizon scales,
before they re-enter the horizon during the late radiation-dominated
era. Motivated by the work of Ref.~\cite{Ota:2023iyh}, we examined whether
this remains true when we account for the fluctuating nature of thermal 
plasma driving the expansion. We consider a simplified model for the 
plasma that is made up of free streaming photons in a thermal state.
We have computed a perturbative one-loop correction to the superhorizon tensor 
power spectrum, showing a multiplicative growing secular 
correction~(\ref{DeltaPtB+C}) that is unsuppressed by any small parameter.
The absence of a suppression factor is a consequence of the Friedmann 
equation~(\ref{Fri})
marrying the physical scales that multiply the one-loop correction,~$\pi^2\kappa^2 T_{\rm rh}^4/(90 H_{\rm rh}^2)\!=\!1$. Despite the weakness 
of the interaction between individual photons in the plasma and the tensor 
modes, the occupation number of photons is high enough to overcome the 
suppression.

Despite the photon's having two physical polarizations,
this enhancement is the same as the one found in Ref.~\cite{Ota:2023iyh},
that considered plasma modeled by a scalar field.
This suggests the amplification might be independent of the number of relativistic degrees of freedom making up the non-interacting plasma. 
Superficially the effect seems additive for the relativistic species, via the factor $g_\text{eff}T_\text{rh}^4$. 
However, it is precisely this combination that is constant for a given model
of inflation and a given reheating mechanism, as evident from the Friedmann equation~(\ref{Fri}), so that the reheating temperature is proportional
to~$g_{\rm eff}^{-1/4}$.

Because of the conformal coupling of photons to gravity we were able to 
treat the plasma at leading order as a thermal
state of photons that is unchanged by the expansion of the Universe apart
from the redshifting of its temperature,~$T\!=\! T_{\rm rh}/a$. One-loop
corrections to the superhorizon tensor perturbations descend from 
interactions with the thermal fluctuations of the photons. The effect
of the local one-loop diagram~\ref{4vertexDiag}
is to ensure that tensor perturbations
are massless at the linearized level, as captured by the Hartree 
approximation, while the enhancement effect derives from the nonlocal
diagrams of Fig.~\ref{diagrams}. The first of these nonlocal diagrams
represents the effect of thermally induced gravitational wave 
production~\cite{Ghiglieri:2015nfa,Ghiglieri:2020mhm}, that
is negligible for observationally relevant scales owing to its blue
tilt and thus suppression on superhorizon scales. It is only the $B$ and $C$ 
type contributions that are relevant here. They represent the effect
of the tensor perturbation disturbing the thermal distribution of
photons, which in turn backreacts back onto the tensor perturbations.
This is seen from the fact that these diagrams arise from the perturbation 
of the Maxwell equation induced by tensor modes. Similar mechanism at one-loop
was found to be relevant when considering corrections to the tensor 
power spectrum during inflation coming from
an excited state of scalar fluctuations~\cite{Ota:2022xni} 
(however, see~\cite{Ema:2025ftj} for a recent criticism of this work.)

We were also able to explore the 
limitations of the late-time one-loop corrected linearized equation for tensor 
perturbations. By solving Eq.~(\ref{MassiveEq}) we have resummed an 
infinite number of self-energy insertions depicted in parts~$B$ and~$C$ 
of Fig.~\ref{diagrams}.
This resummation shows a growing power-law secular 
correction~(\ref{PowerLawCorrection}), which, if taken at face 
value, would imply an enormous correction to primordial tensor perturbations.
Given that the correction grows in time, the effective
representation fo the effect in terms of shear viscosity~\cite{Weinberg:2003ur}
that applies at early times, is invalidated at late times, and rather 
Eq.~(\ref{MassiveEq}) should be used to capture the effect at late times.
However, we should be very careful when analyzing this result, as it is
clearly very soon driven outside of the range of validity of the 
approximations made. Furthermore, the absence of a suppression factor at
one-loop order implies that some of the higher 1-particle-irreducible
loop corrections might be equally unsuppressed, making the problem essentially 
non-perturbative. In that sense the results we report should be taken as 
an indicator of potentially interesting effects, but not as a prediction.
Further work is necessary in order to understand how
to quantify the effect reliably, and there are several aspects to explore.

One should be aware of the limitations of our description for the plasma: 
we are treating it as thermal state of non-interacting photons, that is thermalized on a homogeneous and isotropic cosmological 
background at time of reheating. In reality, 
the primordial plasma is in addition made up of a number charged relativistic 
baryonic species that interact with the photons. This interaction is able to 
maintain thermal equilibrium efficiently, at least over causal length scales. 
This begs the question whether the interacting plasma in the presence of an 
external tensor fluctuation would re-equilibrate more quickly  than in our 
model where plasma interactions are neglected, leading to dampening of the 
one-loop amplification we have found. One possible way of gaining insight  
into the effects of plasma interactions, while maintaining a degree of 
technical simplicity, would be to consider a conformally coupled but 
interacting model for the plasma. However, addressing the question of 
equilibrating upon being perturbed by superhorizon tensor modes would still 
require methods beyond equilibrium thermal field theory as we are dealing 
with an inherently time-dependent system.

Of course, another relevant question is what happens to scalar 
perturbations during the radiation period.
Depending on this, interpretation of the upper bound for tensor-to-scalar 
ratio might require revisiting, and the way we connect the measured 
quantity to the inflationary ratio might require changing. This would also
call for a reassessment of the exclusions bounds of inflationary models.
However, at this stage, the only thing we can safely claim is that the 
issue of large secular enhancements of tensor modes found originally
in~\cite{Ota:2023iyh}, 
and also obtained here should be taken as pointing to an 
important open problem that needs 
resolving. In any case, whatever is necessary to obtain a reliable result is 
likely to update our understanding of the propagation of 
superhorizon modes through the primordial Universe.

\section*{Acknowledgements}
\label{sec: Acknowledgements}

M.B.F. has been funded by the Deutsche Forschungsgemeinschaft (DFG, German Research Foundation) --- project no. 396692871 within the Emmy Noether grant CA1850/1-1. 
D.G. was supported by project 24-13079S of the Czech Science Foundation 
(GA\v{C}R). I.S. acknowledges support of the European Structural and Investment Funds and the Czech Ministry of Education, Youth and Sports (project FORTE —
CZ.02.01.01/00/22\_008/0004632). The authors also acknowledge networking 
support by the COST Action CA23130 BridgeQG.

\appendix
\section{Loop integrals}
\label{appendix:loopintegral}

The two loop integrals in the vanishing external momentum limit needed in 
Sec.~\ref{sec: Computing diagrams} are given by
\begin{align}
I_1(T,\Delta\eta) ={}&
    \int\! \frac{ d^3q }{ (2\pi)^3 } \,
    \Bigl[
    \widetilde{F} \bigl( \Delta\eta \big| q \bigr)
    \Bigr]^2
    \Bigl[ \mathbb{P}^{ijkl}(\vec{k}) 
        q_i q_j q_k q_l \Bigr]_{\vec{k} \to 0}
        \, ,
\\
I_2(T,\Delta\eta) ={}&
    \int\! \frac{ d^3q }{ (2\pi)^3 } \,
    \widetilde{G}_{\scr \rm R}\bigl( \Delta\eta \big| q \bigr)
    \widetilde{F} \bigl( \Delta\eta \big| q \bigr)
    \Bigl[ \mathbb{P}_{ijkl}(\vec{k}) 
        q^i q^j q^k q^l \Bigr]_{\vec{k} \to 0}
        \, ,
\end{align}
where brevity  in this appendix we omit the subscript on the reheating temperature.
The vanishing momentum limit of the contracted tensor structures is best
expressed in spherical integration coordinates in 
which~$\vec{k} \!\cdot\! \vec{q} = kq \cos(\vartheta)$,
\begin{equation}
\mathbb{P}^{ijkl}(\vec{k}) 
    q_i q_j q_k q_l
    =
    \frac{ q^4 }{2}
    \bigl[ 1 - \cos^2(\vartheta) \bigr]^2
    \, .
\end{equation}
Integrals over both angular coordinates can then be performed,
\begin{align}
I_1 (T,\Delta\eta) ={}&
    \frac{2}{15 \pi^2}
    \int_{0}^{\infty} \! dq \, q^6 \,
    \Bigl[
    \widetilde{F} \bigl( \Delta\eta \big| q \bigr)
    \Bigr]^2
    \, ,
\\
I_2 (T,\Delta\eta) ={}&
    \frac{2}{15 \pi^2}
    \int_{0}^{\infty} \! dq \, q^6 \,
    \widetilde{G}_{\scr \rm R}\bigl( \Delta\eta \big| q \bigr)
    \widetilde{F} \bigl( \Delta\eta \big| q \bigr)
    \, .
\end{align}
In the limit where temperature is much larger than the Hubble rate these
integrals are finite,
\begin{align}
I_1 (T,\Delta\eta)
	\xrightarrow{T \gg H}{}&
	\frac{2}{15 \pi^2} \int_0^{\infty} \! dq \,
	\frac{ q^4 \cos^2(q\Delta\eta) }{ \bigl( e^{q/T } - 1 \bigr) \bigl( 1 - e^{-q/T} \bigr) }
    \, ,
\\
I_2(T,\Delta\eta) \xrightarrow{T \gg H}{}&
    -
    \frac{ \theta(\Delta\eta)  }{15 \pi^2}
    \int_0^\infty \! dq \, \frac{ q^4 \sin( 2q\Delta\eta ) }
        { e^{q/T} - 1 }
    \, .
\end{align}
They can both be written in terms of integral~3.911.2 from the
table of definite integrals from~\cite{Gradshteyn:2007},
\begin{equation}
\mathcal{I}(T,\Delta\eta) 
    = \int_0^\infty \! dq \, \frac{ \sin( 2q\Delta\eta ) }
        { e^{q/T} - 1 }
        =
        \frac{\pi T}{2}
        \biggl[
        {\rm cth} \bigl( 2\pi T \Delta\eta \bigr)
        -
        \frac{1}{2\pi T \Delta\eta }
        \biggr]
        \, .
\end{equation}
For the second integral this is accomplished by simply extracting derivatives,
\begin{equation}
I_2(T,\Delta\eta)
    =
    -
    \frac{ \theta(\Delta\eta)  }{240 \pi^2}
    \Bigl( \frac{\partial}{\partial \Delta\eta} \Bigr)^{\!4}
    \mathcal{I}(T,\Delta\eta) 
    \, .
\label{TotalDerIntegral}
\end{equation}
For the first integral this is only a little more involved. It requires
recognizing that,
\begin{equation}
\frac{1}{ \bigl( e^{q/T} - 1 \bigr) \bigl( 1 - e^{-q/T} \bigr) }
    =
    - T \frac{\partial}{\partial q}
    \frac{1}{ \bigl( e^{q/T} - 1 \bigr) }
    \, ,
\end{equation}
then partially integrating the $q$-derivative,
\begin{equation}
I_1 (T,\Delta\eta) =
    \frac{2T}{15 \pi^2}
    \int_0^{\infty} \! dq \,
    \frac{ q^3 \bigl[ 4 \cos^2(q\Delta\eta)
        - 2 q \Delta\eta \sin(q\Delta\eta) \cos(q\Delta\eta) \bigr] }
        { e^{q/T} \!-\! 1 }
    \, .
\end{equation}
Then after using simple trigonometric 
identities~$2 \sin(\alpha) \cos(\alpha) = \sin(2\alpha)$
and~$2 \cos^2(\alpha) = 1 + \cos(2\alpha)$
it becomes clear how to extract derivatives to arrive at the desired
form,
\begin{align}
I_1 (T,\Delta\eta) ={}&
    \frac{2T}{15 \pi^2}
    \biggl[
    2 T^4 \Gamma(4) \zeta(4)
    -
    \frac{1}{4}
    \Bigl(
    1
    +
    \frac{\Delta\eta}{4}
    \frac{\partial}{\partial \Delta\eta}
    \Bigr)
    \Bigl( \frac{\partial}{\partial \Delta\eta} \Bigr)^{\!3}
    \mathcal{I}(T,\Delta\eta)
        \biggr]
    \, ,
\end{align}
where we recognized the definition of the zeta function
\begin{equation}
\zeta(s) = \frac{1}{\Gamma(s)} \int_{0}^{\infty} \! 
    dz \, \frac{z^{s-1} }{ e^z - 1 }
    \, .
\end{equation}
Finally, we have closed-form expressions for both integrals,
\begin{align}
I_1 (T,\Delta\eta) ={}&
    \frac{ 2 \pi^2 T^5}{15}
    \biggl[
    \frac{2}{15}
    +
    \frac{
    3 \, {\rm sh}(2\pi T \Delta\eta)
        +
        {\rm sh}(6\pi T \Delta\eta)
        }{ {\rm sh}^5(2\pi T \Delta\eta) }
    -
    \frac{
    \pi T \Delta \eta
        \bigl[ 11 {\rm ch} ( 2\pi T \Delta\eta ) 
            + {\rm ch} (6\pi T \Delta\eta) \bigr]
        }{ {\rm sh}^5(2\pi T \Delta\eta) }
    \biggr]
    \, ,
\label{I1solution}
\\
I_2 (T,\Delta\eta) ={}&
    \theta(\Delta\eta)
    \frac{\pi^3T^5}{15} \biggl[
        \frac{3}{8( \pi T \Delta\eta )^5}
        -
        \frac{11 \, {\rm ch} ( 2\pi T \Delta\eta ) 
            + {\rm ch} (6\pi T \Delta\eta) }{ {\rm sh}^5 (2\pi T \Delta\eta) }
        \biggr]
        \, .
\label{I2solution}
\end{align}
%



\begin{thebibliography}{99}

\bibitem{Martin:2013tda}
J.~Martin, C.~Ringeval and V.~Vennin,
``Encyclop\ae{}dia Inflationaris: Opiparous Edition,''
Phys. Dark Univ. \textbf{5-6} (2014), 75-235
[arXiv:1303.3787 [astro-ph.CO]].

\bibitem{Planck:2018jri}
Y.~Akrami \textit{et al.} [Planck],
``Planck 2018 results. X. Constraints on inflation,''
Astron. Astrophys. \textbf{641} (2020), A10
[arXiv:1807.06211 [astro-ph.CO]].

\bibitem{Mukhanov:1990me}
V.~F.~Mukhanov, H.~A.~Feldman and R.~H.~Brandenberger,
``Theory of cosmological perturbations. Part 1. Classical perturbations. Part 2. Quantum theory of perturbations. Part 3. Extensions,''
Phys. Rept. \textbf{215} (1992), 203-333

\bibitem{Ota:2023iyh}
A.~Ota, M.~Sasaki and Y.~Wang,
``One-loop thermal radiation exchange in gravitational wave power spectrum,''
JHEP \textbf{03} (2025), 055
[arXiv:2310.19071 [astro-ph.CO]].

\bibitem{Ota:2024idm}
A.~Ota,
``Cosmological stimulated emission,''
Eur. Phys. J. C \textbf{85} (2025) no.7, 813
[arXiv:2412.20474 [astro-ph.CO]].

\bibitem{Liu:2024utl}
L.~Liu and T.~Prokopec,
``Appearances are deceptive: can graviton have a mass?,''
JHEP \textbf{05} (2025), 191
[arXiv:2407.12657 [hep-th]].

\bibitem{Hu:2008rga}
B.~L.~Hu and E.~Verdaguer,
``Stochastic Gravity: Theory and Applications,''
Living Rev. Rel. \textbf{11} (2008), 3
[arXiv:0802.0658 [gr-qc]].

\bibitem{Hu:2020luk}
B.~L.~Hu and E.~Verdaguer,
``Semiclassical and Stochastic Gravity: Quantum Field Effects on Curved Spacetime,''
(Cambridge University Press, Cambridge, 2020)

\bibitem{Ghiglieri:2015nfa}
J.~Ghiglieri and M.~Laine,
``Gravitational wave background from Standard Model physics: Qualitative features,''
JCAP \textbf{07} (2015), 022
[arXiv:1504.02569 [hep-ph]].

\bibitem{Ghiglieri:2020mhm}
J.~Ghiglieri, G.~Jackson, M.~Laine and Y.~Zhu,
``Gravitational wave background from Standard Model physics: Complete leading order,''
JHEP \textbf{07} (2020), 092
[arXiv:2004.11392 [hep-ph]].

\bibitem{Chernikov:1968zm}
N.~A.~Chernikov and E.~A.~Tagirov,
``Quantum theory of scalar fields in de Sitter space-time,''
Ann. Inst. H. Poincare Phys. Theor. A \textbf{9} (1968), 109.

\bibitem{Bunch:1978yq}
T.~S.~Bunch and P.~C.~W.~Davies,
``Quantum Field Theory in de Sitter Space: Renormalization by Point Splitting,''
Proc. Roy. Soc. Lond. A \textbf{360} (1978), 117-134.

\bibitem{Grishchuk:1975uf}
L.~P.~Grishchuk,
``The Amplification of Gravitational Waves and Creation of Gravitons in the Isotropic Universe,''
Lett. Nuovo Cim. \textbf{12} (1975), 60-64
[erratum: Lett. Nuovo Cim. \textbf{12} (1975), 432]

\bibitem{Starobinsky:1979ty}
A.~A.~Starobinsky,
``Spectrum of relict gravitational radiation and the early state of the universe,''
JETP Lett. \textbf{30} (1979), 682-685

\bibitem{Rubakov:1982df}
V.~A.~Rubakov, M.~V.~Sazhin and A.~V.~Veryaskin,
``Graviton Creation in the Inflationary Universe and the Grand Unification Scale,''
Phys. Lett. B \textbf{115} (1982), 189-192

\bibitem{Abbott:1984fp}
L.~F.~Abbott and M.~B.~Wise,
``Constraints on Generalized Inflationary Cosmologies,''
Nucl. Phys. B \textbf{244} (1984), 541-548

\bibitem{Lyth:2009zz}
D.~H.~Lyth and A.~R.~Liddle,
``The primordial density perturbation: Cosmology, inflation and the origin of structure,'' (Cambridge University Press, Cambridge, 2009)

\bibitem{Lewis:1999bs}
A.~Lewis, A.~Challinor and A.~Lasenby,
``Efficient computation of CMB anisotropies in closed FRW models,''
Astrophys. J. \textbf{538} (2000), 473-476
[arXiv:astro-ph/9911177 [astro-ph]].

\bibitem{Lesgourgues:2011re}
J.~Lesgourgues,
``The Cosmic Linear Anisotropy Solving System (CLASS) I: Overview,''
[arXiv:1104.2932 [astro-ph.IM]].

\bibitem{Weinberg:2003ur}
S.~Weinberg,
``Damping of tensor modes in cosmology,''
Phys. Rev. D \textbf{69} (2004), 023503
[arXiv:astro-ph/0306304 [astro-ph]].

\bibitem{Baym:2017xvh}
G.~Baym, S.~P.~Patil and C.~J.~Pethick,
``Damping of gravitational waves by matter,''
Phys. Rev. D \textbf{96} (2017) no.8, 084033
[arXiv:1707.05192 [gr-qc]].

\bibitem{BICEP:2021xfz}
P.~A.~R.~Ade \textit{et al.} [BICEP and Keck],
``Improved Constraints on Primordial Gravitational Waves using Planck, WMAP, and BICEP/Keck Observations through the 2018 Observing Season,''
Phys. Rev. Lett. \textbf{127} (2021) no.15, 151301
[arXiv:2110.00483 [astro-ph.CO]].


\bibitem{Martin:2014nya}
J.~Martin, C.~Ringeval and V.~Vennin,
``Observing Inflationary Reheating,''
Phys. Rev. Lett. \textbf{114} (2015) no.8, 081303
[arXiv:1410.7958 [astro-ph.CO]].

\bibitem{Bezrukov:2007ep}
F.~L.~Bezrukov and M.~Shaposhnikov,
``The Standard Model Higgs boson as the inflaton,''
Phys. Lett. B \textbf{659} (2008), 703-706
[arXiv:0710.3755 [hep-th]].

\bibitem{Allahverdi:2010xz}
R.~Allahverdi, R.~Brandenberger, F.~Y.~Cyr-Racine and A.~Mazumdar,
``Reheating in Inflationary Cosmology: Theory and Applications,''
Ann. Rev. Nucl. Part. Sci. \textbf{60} (2010), 27-51
[arXiv:1001.2600 [hep-th]].

\bibitem{Gross:2024wkl}
M.~Gross, Y.~Mambrini, E.~Kpatcha, M.~O.~Olea-Romacho and R.~Roshan,
``Gravitational wave production during reheating: From the inflaton to primordial black holes,''
Phys. Rev. D \textbf{111} (2025) no.3, 035020
[arXiv:2411.04189 [hep-ph]].

\bibitem{Caprini:2018mtu}
C.~Caprini and D.~G.~Figueroa,
``Cosmological Backgrounds of Gravitational Waves,''
Class. Quant. Grav. \textbf{35} (2018) no.16, 163001
[arXiv:1801.04268 [astro-ph.CO]].

\bibitem{Meda:2020smb}
P.~Meda, N.~Pinamonti and D.~Siemssen,
``Existence and Uniqueness of Solutions of the Semiclassical Einstein Equation in Cosmological Models,''
Annales Henri Poincare \textbf{22} (2021) no.12, 3965-4015
[arXiv:2007.14665 [math-ph]].

\bibitem{Glavan:2017srd}
D.~Glavan,
``Perturbative reduction of derivative order in EFT,''
JHEP \textbf{02} (2018), 136
[arXiv:1710.01562 [hep-th]].

\bibitem{Glavan:2024cfs}
D.~Glavan, S.~Mukohyama and T.~Zlosnik,
``Removing spurious degrees of freedom from EFT of gravity,''
JCAP \textbf{01} (2025), 111
[arXiv:2409.15989 [gr-qc]].

\bibitem{Brown:1976wc}
L.~S.~Brown,
``Stress Tensor Trace Anomaly in a Gravitational Metric: Scalar Fields,''
Phys. Rev. D \textbf{15} (1977), 1469

\bibitem{Bellac:2011kqa}
M.~L.~Bellac,
``Thermal Field Theory,''
(Cambridge University Press, Cambridge, 2011)

\bibitem{Burgess:2003jk}
C.~P.~Burgess,
``Quantum gravity in everyday life: General relativity as an effective field theory,''
Living Rev. Rel. \textbf{7} (2004), 5-56
[arXiv:gr-qc/0311082 [gr-qc]].

\bibitem{Brown:1977pq}
L.~S.~Brown and J.~P.~Cassidy,
``Stress Tensor Trace Anomaly in a Gravitational Metric: General Theory, Maxwell Field,''
Phys. Rev. D \textbf{15} (1977), 2810

\bibitem{Frob:2025uev}
M.~B.~Fr{\"o}b, D.~Glavan and P.~Meda,
``Measurements in stochastic gravity and thermal variance,''
[arXiv:2506.23193 [gr-qc]].

\bibitem{Glavan:2019uni}
D.~Glavan and G.~Rigopoulos,
``One-loop electromagnetic correlators of SQED in power-law inflation,''
JCAP \textbf{02} (2021), 021
[arXiv:1909.11741 [gr-qc]].

\bibitem{Rebhan:1990yr}
A.~Rebhan,
``Collective phenomena and instabilities of perturbative quantum gravity at nonzero temperature,''
Nucl. Phys. B \textbf{351} (1991), 706-734

\bibitem{Ota:2025yeu}
A.~Ota and Y.~Zhu,
``Graviton stimulated emission in squeezed vacuum states,''
[arXiv:2504.06539 [hep-th]].

\bibitem{Ota:2022xni}
A.~Ota, M.~Sasaki and Y.~Wang,
``One-loop tensor power spectrum from an excited scalar field during inflation,''
Phys. Rev. D \textbf{108} (2023) no.4, 043542
[arXiv:2211.12766 [astro-ph.CO]].

\bibitem{Ema:2025ftj}
Y.~Ema, M.~Hong, R.~Jinno and K.~Mukaida,
``Cancellation of one-loop correction to soft tensor power spectrum,''
[arXiv:2506.15780 [astro-ph.CO]].

\bibitem{Gradshteyn:2007}
I.~S.~Gradshteyn and I.~M.~Ryzhik,
``Table of integrals, series, and products,'' seventh edition,
edited by A.~Jeffrey and D.~Zwillinger,
(Elsevier/Academic Press, Amsterdam, Netherlands, 2007).

\end{thebibliography}
\end{document}